\documentclass[referee,useAMS,usenatbib,usegraphicx]{mn2e}
\usepackage{amssymb}
\usepackage{ulem}
\usepackage{color}
\usepackage{hyperref}

\newcommand{\APh}{A.Ph.}

\def\eg{{\it e.g.,~}}
\def\ie{{\it i.e.,~}}
\def\kms{~{\rm km~s^{-1}}}
\def\cm3{~{\rm cm^{-3}}}
\def\muG{~\mu{\rm G}}

\title [Nonthermal Radiation from Supernova Remnants]
{Nonthermal Radiation from Type Ia Supernova Remnants}

\author[Edmon, Kang, Jones, and Ma]
{Paul P. Edmon,$^{1}$\thanks{E-mail: edmonpp@msi.umn.edu},
Hyesung Kang,$^{2}$\thanks{E-mail: kang@uju.es.pusan.ac.kr},
T.~W.~Jones,$^{1}$\thanks{E-mail: twj@astro.umn.edu},
Renyi Ma,$^{3}$\thanks{E-mail: mry@canopus.cnu.ac.kr},\\
$^{1}$Department of Astronomy, University of Minnesota, Minneapolis, MN 55455, USA\\
$^{2}$Department of Earth Sciences, Pusan National University, Pusan 609-735, Korea\\
$^{3}$Department of Astronomy and Space Sciences, Chungnam National University, 305-764, Daejeon, Korea\\}

\begin{document}
\date{\today}
\pagerange{\pageref{firstpage}--\pageref{lastpage}} \pubyear{2010}

\maketitle
\label{firstpage}

\begin{abstract}
We present calculations of expected continuum emissions from Sedov-Taylor phase Type Ia 
supernova remnants (SNRs),
using the energy spectra of cosmic ray (CR) electrons and protons 
from nonlinear diffusive shock acceleration (DSA) simulations.
A new, general-purpose radiative process code, {\it Cosmicp},
was employed to calculate the radiation expected from 
CR electrons and protons and their secondary products.
These radio, X-ray and gamma-ray emissions are generally
consistent with current observations of Type Ia SNRs.
The emissions from electrons in these models dominate the radio through X-ray bands. 
Decays of $\pi^0$s from $p-p$ collisions mostly dominate 
the gamma-ray range, although
for a hot, low density ISM case ($n_{ISM}=0.003\cm3$), the pion decay
contribution is reduced sufficiently to reveal the inverse Compton
contribution to TeV gamma-rays.
In addition, we present simple scalings for the contributing emission processes 
to allow a crude exploration of model parameter space, enabling these results
to be used more broadly.
We also discuss the radial surface brightness profiles expected
for these model SNRs in the X-ray and gamma-ray bands.
\end{abstract}


\maketitle

\section{Introduction}

Diffusive Shock Acceleration (DSA) at astrophysical shocks has become 
the standard model for production of cosmic rays (CRs) 
\citep[\eg][and references therein]{maldru01}.
The CRs observed at the knee ($\sim 10^{15}$eV) and below are most commonly
explained through DSA in the blast waves
of Galactic supernova remnants (SNRs) \citep{bo78,lc83,be87,drury01}. 
Several indirect arguments, including energetics
and composition, support this explanation \citep[\eg][]{gai05,hill06,ellison07}.  

Direct evidence that
at least some SNRs are capable of accelerating electrons and possibly hadrons
to at least tens of TeV comes from observations of nonthermal X-ray filaments
in some historical SNRs \citep[\eg][]{bamba06,pari06}  
and from TeV $\gamma$-rays observed in several shell SNRs
\citep[\eg][]{funk07,lg07}.  
The nonthermal X-rays are likely to be synchrotron emission from
TeV electrons \citep[\eg][]{uchi07}. The TeV $\gamma$-ray origins, although they obviously require TeV or higher energy charged 
particles, have less certain origins and are sensitive to several model parameters.
Inverse Compton scattering, secondary $\pi^0$ decays
and bremsstrahlung may all contribute, but at levels that depend on
details of the accelerated CR spatial and energy distributions as well
as ambient photon, magnetic field and plasma properties.  However, recent observations by {\it Fermi} of several Type II SNRs indicate that secondary $\pi^0$ decay from proton-proton interactions is the most likely explanation for the $\gamma$-ray emission there, which would observationally confirm the SNR origins of CR protons \citep{abdo10}.

In order to test the SNR origins of Galactic CRs, generally, 
and the accuracy and physical assumptions of
DSA model calculations, specifically, it is important to 
determine from simulations not only the CR distributions,
but also the related emissions, so they can be compared
with current and anticipated observations \citep[\eg][]{be87,hill05}. 
Indeed several model SNR comparisons of this type have been published
\citep[\eg][]{ec05,bkv09,morlino09,zira09}. 
Those demonstrate that observed emissions from some individual SNRs can be explained
by the simulations with reasonable model choices, although they 
do not yet provide unambiguous model tests
nor clear confirmation of the specific origins of TeV emissions. 

At this early stage of our understanding
it is especially useful to explore generally the nonthermal emissions produced by CRs
in common classes of SNR models and how they depend on model parameters.
In this spirit we present here an examination of the nonthermal
radio to $\gamma$-ray emissions for Sedov-Taylor phase Type Ia SNRs,
using a set of time dependent simulations that followed the evolution of spherical blasts,
including nonlinear DSA at quasi-parallel shocks with thermal leakage injection 
of CR protons and electrons out of the bulk plasma.
The pressure of CR protons is included in the gasdynamic equations,
while CR electrons are treated as a test-particle component. 
The principal differences expected in the electron and proton CR populations come from
differences in injection rates from the thermal population and from significant 
synchrotron/Compton radiative losses experienced by the highest energy electrons, 
but not the protons. 
We consider four different SNR models, including a range of uniform external 
environments and blast energies.

The structure of
our paper is as follows.  Section 2 provides a brief outline of the simulations 
that include both CR protons and electrons. 
A description of the code used to calculate nonthermal emissions as well as 
results from our emission
calculations are presented in \S 3, while \S 4 summarizes our conclusions.

\section{Nonlinear DSA simulations of SN Ia Sedov-Taylor remnants}

\subsection{The Spherical CRASH code}

In the simulations discussed here
the evolution of the CR modified SNR shock is followed using a one-dimensional, 
spherically symmetric version of the Cosmic RAy  SHock (CRASH) gasdynamic-CR code.
The basic gasdynamic equations modified to implement nonlinear DSA and 
full code details can be found in \citet{kj06}.
CRASH applies adaptive mesh refinement (AMR) techniques and subgrid shock tracking to obtain high
spatial resolution close to the blast shock, where it is crucial for converged
solution of the coupled gas dynamics and CR transport
equations. The spherical CRASH code also incorporates comoving
coordinates expanding with the SNR blast. 
The momentum-dependent CR distribution is evolved through a time-dependent solution of the
diffusion convection equation (DCE).
CRASH incorporates dynamical backreaction from CRs onto the gas dynamics,
including influence of the CR pressure, $P_c$, and
energy exchanges coming from injection of low energy CRs at the gas
subshock and also dissipation of Alfv\'en waves stimulated by streaming CRs.
We note that Berezhko and collaborators have extensively studied the similar problem,
implementing a different, unconventional numerical method for the gasdynamics that normalizes the 
spatial variable by diffusion length at each momentum value and solves the CR 
kinetic equation iteratively to match the downstream and upstream solutions 
at the subshock \citep[\eg][]{berez94,berez97,bkv02}. 
We find that the results of our simulations using the more conventional
CRASH code are quite consistent with these previous studies.

Since DSA operates on relativistic electrons of a given $p$ in exactly the same way as
it does on protons (i.e., it is exclusively rigidity dependent), 
the pitch-angle-averaged phase space distribution functions for CR proton and electron
components, $f_p(p,r,t)$ and $f_e(p,r,t)$
both obey the DCE,
\begin{equation}
{\partial g\over \partial t}  + (u+u_w) {\partial g \over \partial r}
= {1\over{3r^2}} {\partial \over \partial r} \left[r^2 (u+u_w)\right]
\left( {\partial g\over \partial y} -4g \right)
+ {1 \over r^2}{\partial \over \partial r} \left[r^2 \kappa(r,y) {\partial g \over \partial r}\right] 
+ p {\partial \over {\partial y}} \left( {b\over p^2} g \right).
\label{diffcon}
\end{equation}
Here $g$ represents either $g_p= p^4 f_p$ or $g_e=p^4 f_e$, 
$y=\ln(p)$ and $\kappa(r,y)$ is the spatial diffusion coefficient
in the radial direction \citep{skill75}.
Henceforth, particle momenta, $p$, {\it of both protons and electrons}
are expressed in units of $m_pc$, with $m_p$ the proton mass.

The cooling term is $b(p)=-dp/dt=0$ for protons and 
\begin{eqnarray}
b(p)& =&  
- \frac{m_p}{m_e}\frac{\sigma_Tc}{6\pi}\frac{B_{\rm eff}^2}{m_ec^2} p^2\\
&=& -2.37\times 10^{-14} {\rm s^{-1}} \left({B_{\rm eff} \over 100\muG}\right)^2 p^2 \nonumber
\label{ecool}
\end{eqnarray}
for electrons combining synchrotron and IC cooling, 
where $\sigma_T$ is the Thomson cross section,
and $B_{\rm eff}^2= B(r)^2 + B_r^2$ is the effective magnetic field strength
including the energy density of the ambient radiation field. 
We discuss the magnetic and radiation energy densities in \S 3.2.  
The code does not include direct backreaction from large
scale magnetic fields, nor the pondermotive force of the wave turbulence.

The velocity $u_w$ represents the effective radial motion of
scattering centers with respect to the bulk flow velocity, $u$.
Assuming that waves upstream of the gas subshock are predominantly 
resonantly generated through CR streaming and that the shock has a parallel 
magnetic field geometry,
upstream of the gas subshock $u_w$ is set to the Alfv\'en speed, $v_A>0$.
The current version of CRASH does not follow the self-consistent evolution of the
magnetic field strength through wave-particle interactions.
So here we simply set $v_A = B_0/ \sqrt{4\pi \rho(r)}$,
where $B_0$ is the upstream mean magnetic field strength. 
Downstream, the simulations assume $u_w = 0$, since the Alfv\'enic
turbulence in that region is probably relatively balanced.
The transition in scattering center motion, $\Delta u_w>0$, reduces the effective
velocity difference experienced by CRs across the shock
compared to the bulk flow, $\Delta u<0$, thus reducing the DSA efficiency.
Gas heating due to Alfv\'en wave dissipation is represented by the term
\begin{equation}
W(r,t)= - \omega_H v_A {\partial P_c \over \partial r },
\label{wdis}
\end{equation}
where $\omega_H$ is a commonly used dimensionless parameter that controls the degree of dissipation. 
This dissipation term derives from a simple model in which Alfv\'en waves are 
resonantly amplified by streaming CRs in balance with local wave dissipation processes
\citep[\eg][]{jon93}.
As previously shown in SNR simulations \citep[\eg][]{berez97,kj06,capri10,pzs10},
and in more general contexts in earlier work \citep[\eg][]{mckenzi82,mark92,jon93},
Alfv\'enic drift and precursor heating by wave dissipation both reduce
DSA efficiency and associated shock modification.
The significance of Alf\'ven wave dissipation and drift can generally
be characterized in terms of the Alf\'venic Mach number, \ie $v_A = u_0/\mathcal{M}_A$,
where $u_0$ is the upstream plasma flow speed relative to the shock. In these
simulations $\mathcal{M}_A \gg 1$, so the influence is modest.

Bohm-like spatial CR diffusion in the radial direction was used for both protons and electrons
in the simulations; namely,
\begin{equation}
\kappa(r,p) = \kappa_n \cdot p \cdot \left ( \frac{\rho_0}{\rho(r)} \right ) , 
\label{kappa}
\end{equation}
where $\kappa_n = m_pc^3/(3eB_0) = 3.13\times 10^{22}~{\rm cm^2 s^{-1} B^{-1}_\mu}$, 
$B_{\mu}$ is the upstream magnetic field strength in micro-Gauss, 
and $\rho_0$ is the upstream density. 
The density dependence in this
diffusion model represents enhancement of resonant Alfv\'en wave
amplitudes through compression.
 
Low energy CR protons and electrons were injected at the shock in the same manner
through thermal leakage \citep{gies00,kjg02}. 
In this model, thermal ions in a Maxwellian distribution immediately downstream of the
gas shock have a finite probability to escape upstream into the low energy CR
population, 
provided they have sufficient upstream-directed momentum
to overcome transverse MHD waves generated through the cyclotron instability
in the shock transition \citep{malvol98}.
This behavior is modeled numerically through a ``transparency function''.
The function contains
one adjustable parameter, $\epsilon_B=B_0/B_{\perp}$, which compares the
upstream magnetic field, $B_0$, with the amplitude of postshock, amplified MHD waves
that interact with low energy particles, $B_{\perp}$. 
The value $\epsilon_B=B_0/B_{\perp}=0.25$ was used for the models considered here. 
The resulting fraction of thermal protons
injected into the CR pool, $\xi \sim 10^{-3}$. 
It was shown previously that the DSA efficiency saturates at such a high 
injection rate (if $\xi > 10^{-4}$ for strong shocks), 
so except for details near start up, the SNR simulation results 
are insensitive to $\epsilon_B$. The temperature used for the thermal population was computed from the gas pressure, $P_g$, and density, $\rho$,
as $T = \mu m_p P_g/(k_B \rho)$ with $\mu = 0.61$, so assuming equal proton and electron temperatures.

Because postshock thermal electrons have smaller gyroradii, compared to 
thermal protons, the injection rate of electrons should be much smaller.
The ratio of CR electron to proton number injected at the shock,
$K_{e/p} \sim 0.01$, is commonly assumed, 
since about 1\% of the observed Galactic CR flux near a GeV is 
due to electrons \citep{sch02,reynolds08}. 
On the other hand, \citet{morlino09} point out that the CR electron flux observed here at Earth is actually a 
convolution of CR electrons from SNR and other electron accelerators of all ages.  
It is also entirely possible that CR electron injection, which is not well understood, and 
acceleration efficiency change dramatically for SNRs of different ages.   
On the other hand, it is commonly found in models of young SNRs that the best fit to the observational data requires
$K_{e/p} \sim 10^{-4}$ \citep[\eg][]{bkv09,morlino09,zira09},
so we set $K_{e/p} = 10^{-4}$ in our calculations.  However, since the electron
population is passive and the SNRs are optically thin in emission bands of relevance here, our resulting emissions can be generally linearly
scaled to another preferred value of $K_{e/p}$.

In flows where CR backreaction is important, the formation of a CR precursor compresses and heats the inflowing plasma.
These developments can lead to substantial changes in the strength of the
dissipative, gas subshock and to the postshock conditions relative to
those in a pure gas dynamic shock of the same Mach number. Given
these flow modifications in front of the subshock, it is helpful
in the following discussion to identify specifically
the unmodified, upstream conditions by the subscript '0', the conditions
immediately upstream of the gas subshock by the subscript '1', and
the conditions immediately downstream of the full shock structure by the
subscript '2'.

\subsection{Remnant model parameters}

\begin{table*}
 \centering
 \begin{minipage}{140mm}
  \caption{SNR model parameters}
\begin{tabular}{ccccccccc}
\hline
\hline
Model & $n_{ISM}$ & $T_{ISM}$ & $E_o$ & $B_0$  & $r_o$ & $t_o$ &
$u_o$\\
 & ($\rm cm^{-3}$) & (K) & ($10^{51} \rm ergs$) & ($\rm \mu G$)  & (pc) & (years) &
($10^4$ km/s)\\
\hline
S1 & $0.3$ & $3\times 10^4$ & $1$ & $30$  & $3.19$ & $255.$ & $1.22$\\
S2 & $0.3$ & $3\times 10^4$ & $4$ & $30$ & $3.19$ & $127.$ & $2.45$\\
S3 & $0.3$ & $3\times 10^4$ & $1$ & $5$  & $3.19$ & $255.$ & $1.22$\\
S4 & $0.003$ & $10^6$ & $1$ & $5$ & $14.8$ & $1182.$ & $1.22$\\
\hline
\end{tabular}
\\
\textbf{Note: The model ISM Alfv\'en speed, $v_A = 101~\rm{km/s}$ for S1,S2; 
$v_A = 16.8~\rm{km/s}$ for S3; $v_A = 168~\rm{km/s}$ for S4.}
\end{minipage}
\end{table*}

Table 1 lists the dynamical parameters for the SNR models: 
the uniform ambient density, $n_{ISM}$, the ISM temperature, $T_{ISM}$, 
the upstream magnetic field strength, $B_0$, 
and the SN explosion energy $E_o$.
The parameters for S1- S3 (S4) represent warm (hot) phase ISM environments.
Models S1, S3, and S4 assume an explosion energy, $E_o = 10^{51}$erg, while
model S2 begins with a blast containing four times this energy, so $E_o = 4\times 10^{51}$erg.
All models assume there are no pre-existing CRs in the ambient ISM.

The SN ejecta mass is set to $M_{ej} = 1.4M_{\odot}$. 
The simulations are intended to follow evolution during the
adiabatic, Sedov-Taylor (ST) evolution stage. 
Thus, convenient normalization variables are 
$\rho_o=2.34\times 10^{-24}n_{ISM}{\rm ~g ~cm^{-3}}$ for density,
along with ST similarity scales
$r_o = \left[3 M_{ej}/(4\pi \rho_o)\right]^{1/3}$ for length, 
and $t_o = (\rho_o r_o^5/E_o)^{1/2} = [(3 M_{ej}/(4\pi)]^{5/6}\rho_o^{-1/3} E_o^{-1/2}$ for time.
The velocity and pressure scales are defined as $u_o=r_o/t_o = \sqrt{4\pi E_o/(3 M_{ej})}$ 
and $P_o = \rho_o u^2_o = E_o/r_o^3$, respectively.
{\it{In order to avoid confusion later, we note here a distinction between ST
normalization subscripts, '$o$' and initial or upstream condition subscripts '$0$'.}}

It is worth noting that the mass swept up by the blast at time $t$
and radius $r_s=\xi_s r_o (t/t_o)^{2/5}$ is
given in the ST solution as

\begin{equation} 
M_{sw}(t) = M_{ej}\xi_s^3 (r_s/r_o)^3 =  M_{ej}\xi_s^3 (t/t_o)^{6/5},
\end{equation} 

where $\xi_s \approx 1.15$ is the ST similarity constant for a blast in a uniform medium with 
the gas adiabatic index $\gamma_g = 5/3$.
The ST shock speed can be expressed as 

\begin{equation}
U_{ST} = u_0 = \frac{2}{5}\xi_s u_o \left ( \frac{t}{t_o} \right )^{-3/5}= \frac{2}{5} \xi_s^{5/2} u_o \left ( \frac{M_{ej}}{M_{sw}} \right )^{1/2} = 4.6\times 10^3 \kms \left (\frac{u_o}{10^4\kms} \right ) \left (\frac{t}{t_o} \right)^{-3/5} .
\end{equation}

The second to last relation reveals that $U_{ST} \propto \sqrt{E_o/M_{sw}}$,
emphasizing that the dynamical state of a ST blast is determined
by the blast energy and the mass contained within the blast.
Thus, except for numerical factors of order unity, many of the
results we present below can be approximately represented independent of
the ambient radial density distribution in terms of $M_{sw}$
in combination with $E_o$ and $M_{ej}$ rather than $t/t_o$ and $u_o$.

Recent X-ray observations of young SNRs reveal in many cases magnetic fields 
stronger by at least an order of magnitude than the average ISM field 
\citep[\eg][]{bamba06,pari06}.
The existence of such strong fields is now commonly interpreted as the result of
amplification within the shock precursor, either through resonant \citep{lucek00},
or nonresonant \citep{bell04} streaming instability, or possibly hydrodynamic
instability driven by the CR pressure gradient in the precursor \citep{bere09}.
The simulations discussed here do not follow self-consistent amplification in the precursor of the 
magnetic field strength through these processes.
Instead, to provide field values in the model SNRs consistent with
observations, the upstream magnetic field strengths in SNR models S1 and S2 
were set to the relatively large value, $B_0=30 \mu$G.
For comparison, the S3 and S4 models adopted, $B_0 = 5 \mu G$, which is
similar to the mean ISM magnetic field.

In the simulation of the electron synchrotron energy losses
and the resulting synchrotron emission, we assume that 
the field strength is modified during passage through the shock and into the 
SNR interior in a way that maintains consistency with the
diffusion coefficient model (equation [\ref{kappa}]). 
This $\kappa$ model assumes in the relativistic limit
that the scattering length, $\lambda \propto p/(\rho B_0)$, 
along with a fixed ratio of wave field to total field strength. Consequently,
it assumes $B \propto \rho$, so that gas compression through the modified
shock, $\chi = \rho_2/\rho_0 \sim 4 - 13$ (see Fig. 2), leads to postshock field
values in these models, $B_2  = \chi B_0 \sim 120 - 400 ~\mu$G in S1 and S2 and $\sim 20 - 50 ~\mu$G in S3 and S4.

In all the simulations CR backreaction quickly increases 
compression from the initial $\chi = 4$ to $\chi \sim 7~-~13$. Subsequently,
$\chi$ decreases slowly in each case roughly as $\chi \propto t^{-k}$, with 
$k \sim 1/7~-~1/5$ (see Fig. 2). 
\citet{krj09} found for strong, CR modified, plane shocks
that approximately $\chi \propto \mathcal{M }^{1/3}$, where $\mathcal{M}$ is the shock sonic Mach
number. For these SNR simulations, $\mathcal{M}\propto t^{-2/5}$, so
their result would predict $k \sim 2/15$, which is reasonably consistent. 
For ST SNRs one has more generally that $\mathcal{M}\propto M_{sw}^{-1/2}$, so
for scaling relations discussed in \S 3 we will assume that $\chi \propto M_{sw}^{-1/6}$.

The SNR simulations start with ST similarity blast waves at a time $t/t_o = 0.5$,
because the detailed evolution before that time does not strongly affect the later development.
In fact, hydrodynamic simulations of SN blast waves (without the CR terms) 
show that the evolution of the outer shock speed can be approximated by $U_{ST}$
for $t/t_o > 0.2$, although the true ST solution is established only after the 
inner reverse shock is reflected at the center at $t/t_o \sim 7$,
\citep[\eg][]{kang06}.  The simulations are carried out to $t/t_o = 15$, which
is deemed sufficient to establish basic ST phase properties of
the CR population and their dynamical impact until either
the blast strength weakens substantially or the blast becomes radiative.
Earlier studies showed that the highest momentum produced by DSA
during the expansion of an SNR, $p_{\rm max}$, is achieved near the end of the
free expansion stage, so when $t/t_o \sim 1$, and the transfer of explosion
energy to the CR component occurs mostly during the early ST stage, so also when
$t/t_o \sim 1$ \citep[\eg][]{berez97}.

\subsection{Evolution of the CR modified SNRs}

\begin{figure}
   \centering
   \includegraphics[width=0.90\textwidth]{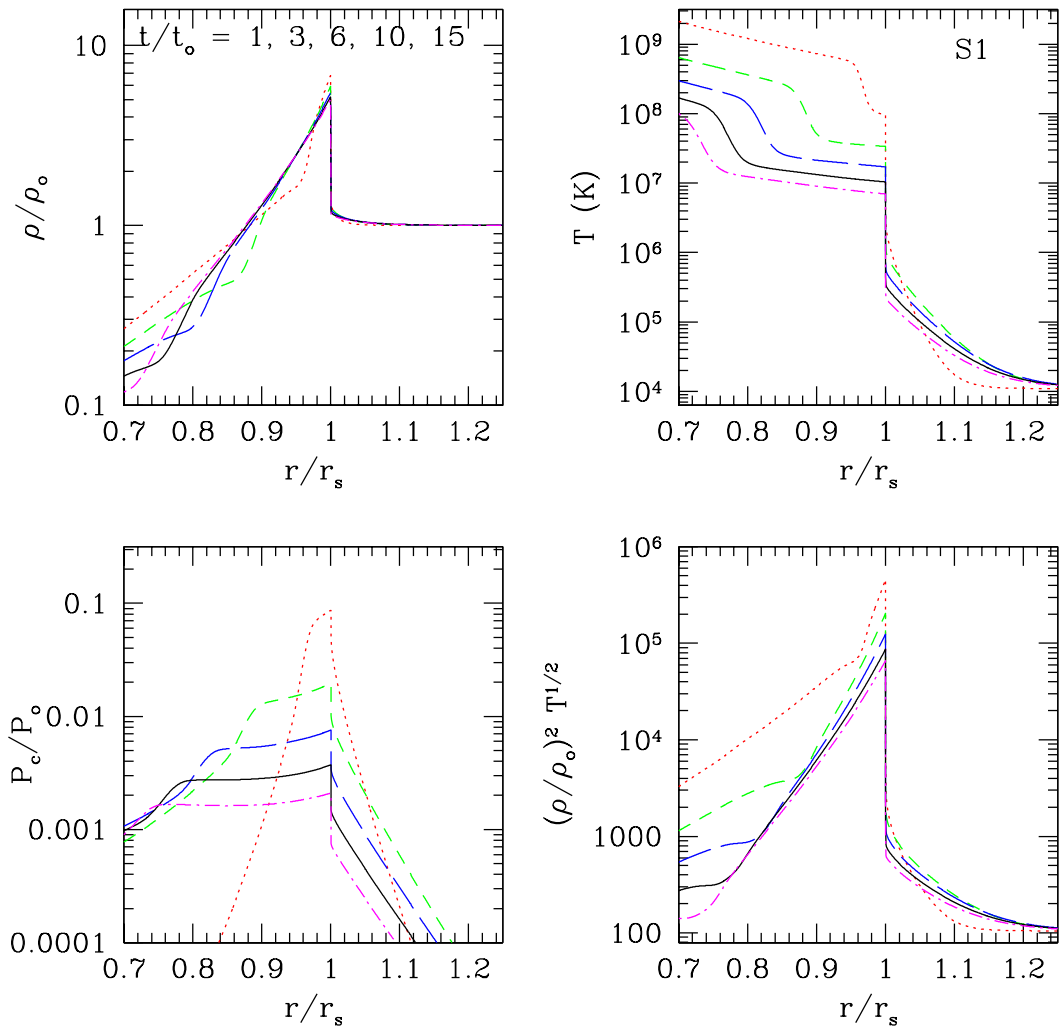}
  \caption{
The time evolution of gas density and temperature for model
S1 along with CR pressure and the factor $\rho^2 T^{1/2}$, which provides
a scaling for thermal bremsstrahlung. Times shown are: $t/t_o = 1$ (red dotted), 
3 (green short dashed),  6 (blue long dashed),  10 (black solid), and 15 (magenta dot-dashed).
(A colour version of this figure is available in the online journal.)}
\end{figure}

We now summarize the distinctive time evolutionary features
of the SNR models used in this study, focusing on nonlinear DSA influences
that impact radiative emissions.
Fig. 1 shows radial profiles from model S1 of several useful 
quantities near the shock at times $t/t_o = 1, 3, 6, 10, 15$. The S1 model 
is representative in this regard of all the simulations. 
Gas density and temperature are shown in the top two panels. They are combined
in the lower right to show $\rho^2 T^{1/2}$ as a proxy for
bolometric thermal bremsstrahlung emissivity. The CR pressure profile is
included in the lower left panel.
To simplify the plots, initial ST conditions from $t/t_o = 0.5$ are not included.  
At the first time shown in Fig. 1, $t/t_o = 1$, the shock
is modified substantially through compression and heating in the CR precursor. 
In fact, the CR acceleration and nonlinear modification peak early in the evolution
and then decrease in time as the shock slows down (see also Fig. 2).  

CR modification of the shock enhanced the postshock density but reduced the 
gas pressure, leading to a postshock gas temperature
lower than that expected for the ST shock near $t/t_o \sim 1$.
The effect of CR pressure
on the temperature distribution behind the shock can be seen by noting that 
the slope of temperature rise towards the interior steepens as
$P_c(r)$ begins to drop sharply inside the blast. 
This interior drop in $P_c(r)$ is
an artifact of the location of the contact discontinuity separating matter
interior and exterior to the initial conditions shock. Its shape depends 
on the rate of spatial CR diffusion into the blast cavity.

In addition to raising the total density jump within the shock, compression
through the precursor also preheats gas before it enters the subshock structure.
This adiabatic effect by itself would raise the upstream temperature by 
a factor $T_1/T_0 \sim (\rho_1/\rho_0)^{2/3}$, or about 1.6 in the S1 model. 
In fact, Fig. 1 shows that in the S1 model $T_1/T_0 \approx 10-100$. 
The additional heating comes from MHD wave dissipation (equation [\ref{wdis}]). 
Compression through the precursor also decelerates upstream flow before it enters
the subshock. All these effects greatly weaken the subshock.
So the sonic Mach number of the subshock decreases quickly to $\mathcal{M}_{sub} \approx 4.8$ 
in the S1 and S4 models and to $\mathcal{M}_{sub} \approx 3.7$ in S3 model by $t/t_o\approx 0.7$ 
and remains approximately the same afterward,
even though the total shock sonic Mach number decreases roughly 
as $M\propto (t/t_o)^{-3/5}$. 
Since $\mathcal{M}_{sub}$ stays constant throughout the evolution, 
the CR injection fraction ($\xi$) via thermal leakage also stays constant 
after $t/t_o \approx  0.7$ (see Fig. 2).
With the adopted injection parameter ($\epsilon_B=0.25$), the CR injection fraction is
high, $\xi\approx 10^{-3}$, and so the CR acceleration should be close to maximum 
efficiency.

These shock modifications also significantly influence the thermal
bremsstrahlung emission in the SNR, whose distribution 
is represented by the $\rho^2 T^{1/2}$ profile in the lower right panel of 
Fig. 1.   The
thermal bremsstrahlung shell will appear thinner in the
strongly modified SNRs than in the analogous gas dynamic SNR. 
The enhanced postshock density also enhances the thermal bremsstrahlung 
emissivity. 
The thinner shell and higher emissivity in the
shell combine for a relatively small difference in the bolometric thermal
bremsstrahlung luminosity compared to an unmodified shock.
The thermal bremsstrahlung luminosity in a fixed spectral band below
the spectral cutoff should generally be enhanced, however, since that
has an approximate $1/T$ dependence with respect to the bolometric emission. 
Of course, the lowered temperatures behind the modified shock
also move the upper cutoff of the thermal bremsstrahlung SED from the hard
X-ray band into the soft X-ray band. 

\begin{figure}
   \centering
   \includegraphics[width=0.90\textwidth]{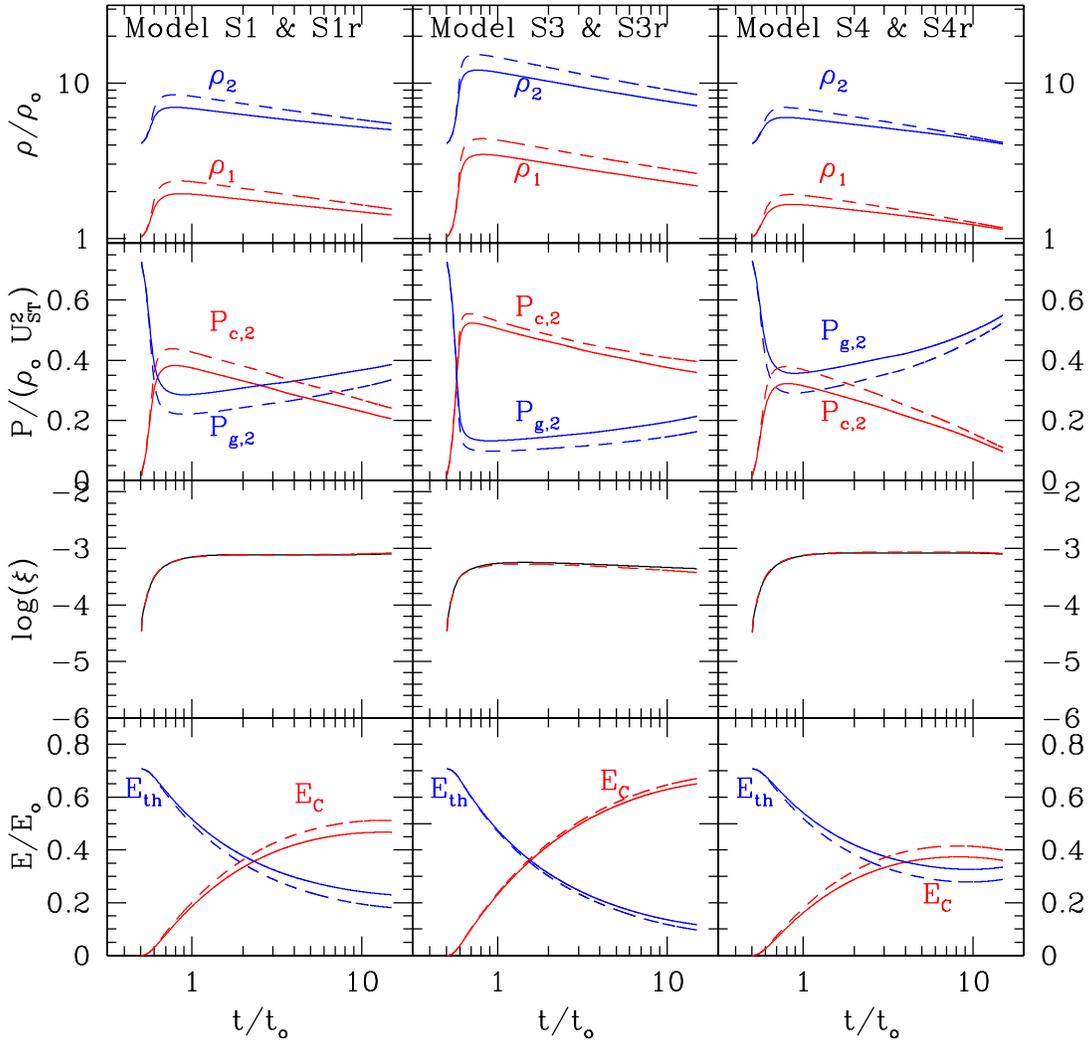}
  \caption{
The time evolution of the compression factors, $\rho_1/\rho_0$
and $\chi=\rho_2/\rho_0$, postshock pressures, $P_{g,2}$ and $P_{c,2}$,
the injection fraction, $\xi$, and the volume integrated energy,
$E_{th}$ and $E_{c}$ is shown for six models.
The wave heating parameter is $\omega_H=0.5$ for S1r, S3r and S4r models,
while $\omega_H=1.$ for S1, S3, and S4 models.  The r models are the dashed lines. (A colour version of this figure is available in the online journal.)
}
\end{figure}

More insights to the evolutionary behaviors of the S1, S3 and S4 models can be
obtained from Fig. 2. As already noted, both the precursor compression, $\rho_1/\rho_0$, and 
the total shock compression, $\chi =\rho_2/\rho_0$, as well as the CR pressure
relative to $\rho_0 U_{ST}^2$ increase quickly from their initial values
\footnote{
The very early evolution is affected by numerical start-up of
the simulations.},
reach their maximum values around $t/t_o \approx 0.7$, and
decrease slowly in time as the shock slows down.
In case of the S2 model with higher $E_o$ (not shown in Fig. 2), the CR acceleration
and the nonlinear modification are slightly more significant compared to the S1 model,
because of a faster shock speed.
If the Alfv\'enic drift and heating were not included, 
the temperature of the ambient medium (and so the sonic Mach number) 
would be the primary parameter that determines the CR injection and 
acceleration efficiencies. 
Compare, for instance, the S3 model in warm ISM to the S4 model in the hot ISM.
Because of the ISM temperature differences, the sonic Mach number of
the total shock in the S3 model is $6$ times higher than S4 at a given $t/t_o$.
So the S3 SNR would be the more efficient CR accelerator. 
The low density and high temperature ISM of the S4 model 
produce low sonic and
Alfv\'enic Mach numbers ($\mathcal{M} \propto T^{-1/2}$, $\mathcal{M}_A=u_s/v_A \propto {n_{ISM}}^{1/2}/B$), 
making
the CR acceleration much less efficient in that model.

The magnetic field strength also has a significant influence on
the SNR evolution even though no direct MHD dynamical effects are modeled. 
Most important to these simulations, the field strength sets the CR spatial diffusion
rate and thus, in concert with $u_o$, the CR acceleration rate (see \S 2.4, below).
Consequently, particles are accelerated to several times higher 
momenta ($p_{\rm max}$) in the S1 model in comparison to the S3 model because of 
the enhanced magnetic field and smaller diffusion coefficient (see Fig. 3 below). 
On the other hand, the stronger field leads to a faster Alfv\'en speed, 
which tends to reduce DSA efficiency by adding entropy to the precursor. 
This reduces the effective velocity jump across the subshock 
and also reduces the subshock Mach number. 
Those effects are responsible for the moderate reductions in CR acceleration 
efficiency in the S1 model compared to the S3 model. 
The effect of Alfv\'en wave heating is further illustrated in Fig. 2, 
where we show three additional simulations, S1r, S3r, and S4r, in which
the wave dissipation parameter is reduced from $\omega_H = 1$
to $\omega_H=0.5$. The smaller wave dissipation rate reduces
non-adiabatic heating in the precursor, allowing greater
compression through the precursor, so an increase in CR acceleration efficiency.

Fig. 2  also shows the volume integrated thermal energy 
($E_{th} = \int 4 \pi r^2 dr (3/2)P_g$)
and CR energy in units of the SN explosion energy.
The energy transfer to CRs seems to saturate for $t/t_o\ga10$,
and the fraction of the blast energy transferred to CRs ($E_c/E_o$) is 
about 0.7, 0.5, 0.45, 0.35 in the S3, S2, S1, and S4 models, respectively.

In summary, adopting simple models for Alfv\'en wave transport,
these simulations demonstrate that the CR acceleration would be less 
efficient at SNRs in hot rarefied ISM where
both the sound speed and Alfv\'en speed are faster, compared to
those in a warm phase ISM.

We note that the evolution of CR modified SNRs shown in Fig. 2 
(especially, $\rho_1$, $\rho_2$, $P_{g,2}$ and $P_{c,2}$) differ somewhat from
the similar SNR models presented in \citet{kang06}.
Several parameters were different there,
\eg $\mu=1$, $\epsilon_B=0.16-0.2$, $T_{ISM}=10^4$K for S1-3 models,
and $B_0=30\mu$G for the S4 model.
The main differences result, however, from a coding error in those earlier
simulations, 
which caused the Alfv\'enic drift and heating effects to not be fully implemented. 
The results in the present study are, however, consistent with independent
simulations of similar SNRs \citep{bkv02},
in which a Type Ia SNR with $E_o=3\times10^{51}$erg and $\xi=2\times10^{-4}$ 
in warm phase ISM with $n_{ISM}=0.3\cm3$, $T_{ISM}=10^4$K, 
and $B_0=20\mu$G was calculated. 

\subsection{The CR proton and electron populations}

\begin{figure}
   \centering
   \includegraphics[width=0.90\textwidth]{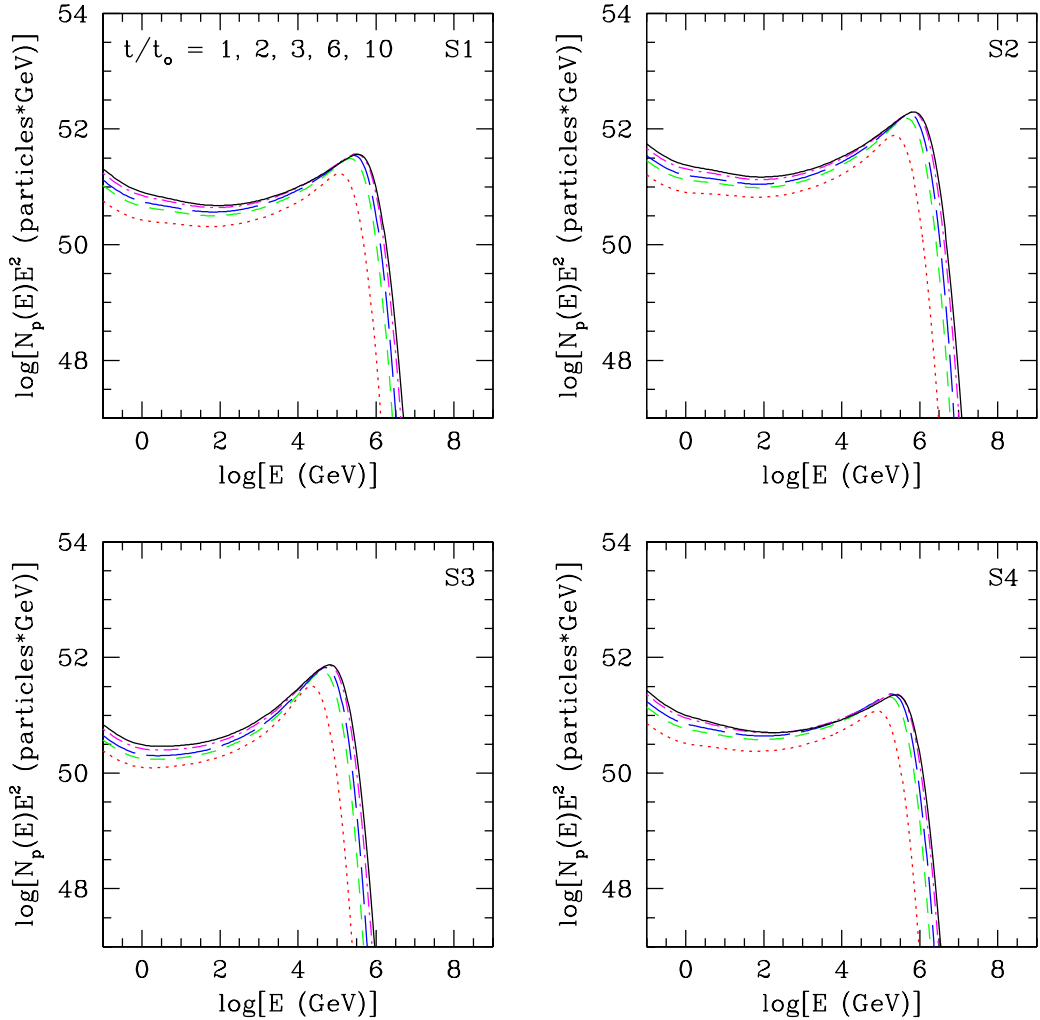}
  \caption{
Volume integrated proton spectra in terms of particle kinetic energy
at $t/t_o = 1$ (red dotted), $2$ (green short dashed), $3$ (blue long dashed),
$6$ (magenta dot-dashed), and $10$ (black solid). (A colour version of this figure is available in the online journal.)
}
\end{figure}

Fig. 3 shows the volume integrated CR proton energy spectrum,
$E^2\cdot N_p(E,t) = \int 4 \pi r^2 dr E^2 n_p(E,r,t)$,
where $E = m_p c^2 (\sqrt{1 + p^2} - 1)$, is the proton kinetic energy,
and $n_p(E,r,t)dE = 4\pi p^2 f(p) dp=n(p,r,t)dp$.
We note that in the models with enhanced $B_0$ (\ie S1 and S2) particles of charge Z can be 
accelerated to $E_{\rm max} \approx Z \cdot 10^{15}$eV
by the early ST stage.
This is consistent with results of other analogous SNR simulations \citep[\eg][]{berez07}.
In the early ST stage the shock structure is significantly
modified, with $\chi = \rho_2/\rho_0\sim 7-10$, so $E^2\cdot N(E)$ is concave upwards. 
This well-known behavior is a consequence of momentum dependent diffusion
across the precursor; 
CRs near the injection momentum experience only the subshock velocity jump, 
while higher momentum CRs see a larger velocity jump.
Thus, the CR spectrum is softer at lower momentum
and harder just below the high energy cutoff 
than the strong shock test particle result, $N(E) \propto E^{-2}$.
Moreover, the Alfv\'enic drift in the precursor further softens the spectrum
of newly accelerated CRs as the shock slows down and the Alfv\'enic Mach number decreases.

It will be useful below to estimate the maximum CR proton momentum expected at a given
time $t/t_o$. This we do by integrating the standard expression for
the average momentum gained per unit time for a particle \citep{lc83},

\begin{eqnarray}
\frac{dp}{dt}& =& \frac{u_0(1 - \mathcal{M}_A^{-1})-u_2}{3} \left ( \frac{\kappa_0}{u_0(1-\mathcal{M}_A^{-1})} + \frac{\kappa_2}{u_2} \right )^{-1} p, \\
 &= &\frac{[u_0(1 - \mathcal{M}_A^{-1})]^2}{q_0\kappa_0} 
\frac{p}{(1+\frac{\kappa_2}{\kappa_0}\frac{q_0}{q_0-3})} , \nonumber
\label{dpacc}
\end{eqnarray}
where, once again, subscripts $0$ and $2$
refer to upstream and downstream conditions in the shock frame, respectively.
The useful parameter, 
\begin{equation}
q_0 =3\frac{u_0-v_A}{u_0-v_A-u_2}=3\frac{\chi(1-\mathcal{M}_A^{-1})}
{\chi(1-\mathcal{M}_A^{-1})-1}
\approx \frac{3\chi}{\chi-1},
\label{q0def}
\end{equation}
is the DSA test particle spectral index, including 
the upstream Alfv\'en wave drift term, $|u_w| = v_A =u_0/\mathcal{M}_A$,
for consistency with equation (\ref{diffcon}). The right-most expression for $q_0$
neglects this term, since it is a relatively minor correction in our model SNRs.

Neglecting the Alf\'ven wave term and assuming a constant compression ratio, $\chi$,
but including the diminishing shock speed in the ST solution,
equation (\ref{dpacc}) gives for protons accelerated between $t_i$ and $t$, 
\begin{equation}
p_{\rm max} \approx \frac{0.54 u_o^2 t_o}{ \kappa_n}~\frac{1}{q_0}
\left(\frac{t_o}{t_i}\right)^{0.2}\left[ 1 - \left(\frac{t_i}{t}\right)^{0.2}\right].
\label{pmax}
\end{equation}
For our simulations, which start from $t_i/t_o = 0.5$, 
equation (\ref{pmax}) asymptotes to $p_{\rm max} \approx 0.15(t_o u_o^2/\kappa_n)
\approx 3.8\times 10^{13}(E_o/10^{51})^{1/2} B_{\mu} n_{ISM}^{-1/3}~$eV/c at large $t/t_o$
using $\chi = 4$ for convenience.
At $t/t_o=10$ this gives, for example, 
$p_{\rm max} \approx 7.8\times 10^{14}$ eV/c for the S1 model
and  $p_{\rm max} \approx 5.8\times 10^{14}$ eV/c for the S4 model.
These estimates are consistent with the numerically determined cutoff energies in $E^2\cdot N_p(E)$
shown in Fig. 3. 

\begin{figure}
   \centering
   \includegraphics[width=0.90\textwidth]{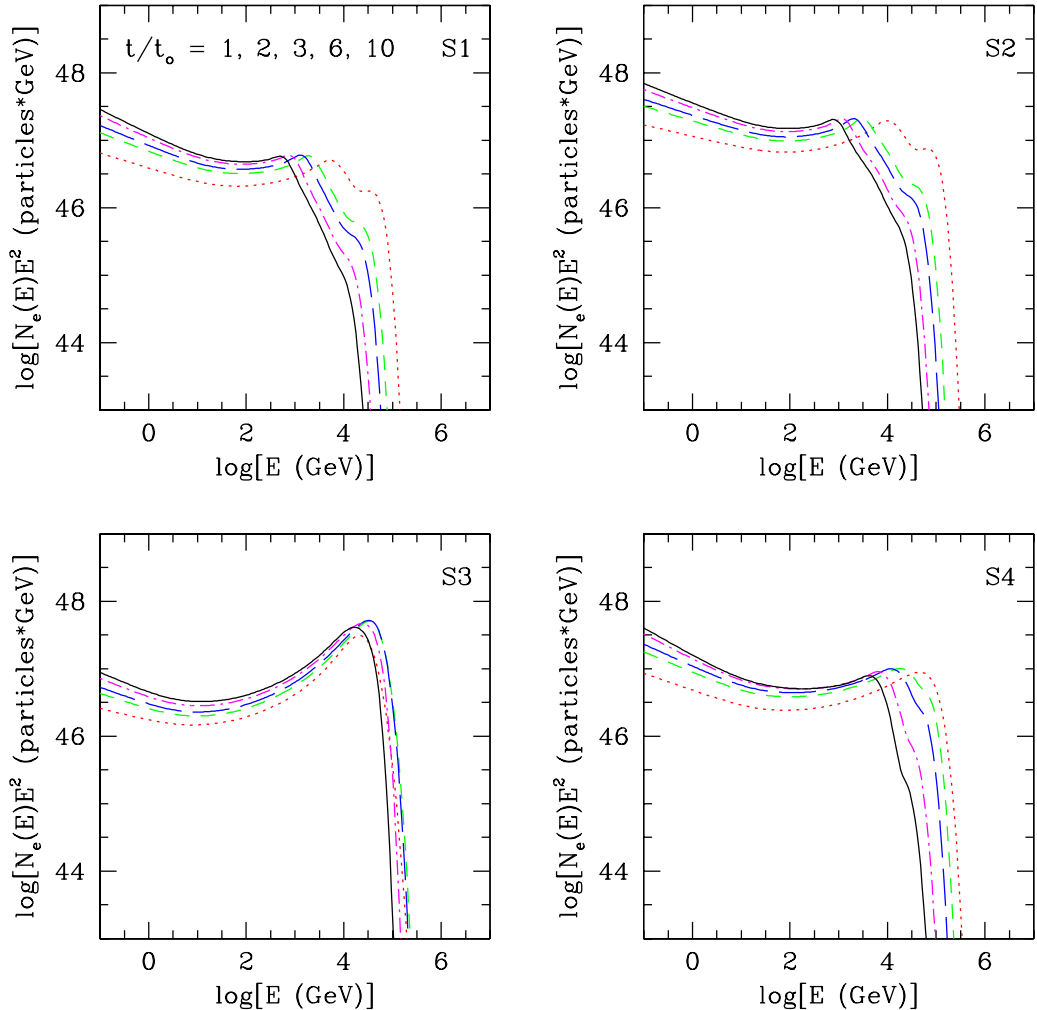}
  \caption{
Volume integrated electron spectra in terms of particle kinetic energy
at $t/t_o = 1$ (red dotted), $2$ (green short dashed), $3$ (blue long dashed),
$6$ (magenta dot-dashed), and $10$ (black solid). (A colour version of this figure is available in the online journal.)
}
\label{necint}
\end{figure}

The CR electron spectra develop substantial differences from the proton
spectra, despite their common DSA interactions \citep[\eg][]{webb84,bkv09,blas10,zira09}.
Fig. 4 shows the volume integrated CR electron energy spectra,
$E^2\cdot N_e(E,t) = 4\pi \int r^2 E^2\cdot n_e(E,r,t)dr$, at several times
for each SNR model.  
Note that the electron spectra cutoff at much lower energies than the analogous proton spectra shown 
in Fig. 3.
The structures of the integrated electron spectra are complex around and below the cutoff. 
The low energy break in each spectrum identifies the energy at which
the synchrotron/Compton loss time for electrons downstream of the shock equals the SNR age. 
Below this energy radiative energy losses are negligible.
The integrated spectrum between this break and the high energy cutoff is 
controlled by the downward movement of the electron energy cutoff as recently accelerated electrons move into the SNR interior away from the shock.  The high energy cutoff itself develops at the shock and corresponds to the energy where synchrotron losses balance DSA gains.

At the shock the electron cutoff momentum, $p_{ec}$,
can be evaluated by balancing DSA and radiative losses. 
In particular, $p_{ec}$ corresponds to the momentum at which the
average momentum gain by an electron in one pair of shock crossings, 
$\Delta p_{DSA} = (4/3)p[u_0(1 -\mathcal{M}_A^{-1}) - u_2]/v$, equals the 
momentum loss from radiation during the same period of time, 
$\Delta p_r = b_2 4 \kappa_2/[u_2v]+b_0 4 \kappa_0/[(u_0(1-\mathcal{M}_A^{-1})v]$, 
where $v=c\sqrt{1-(m_e/(m_p p))^2}$ is the electron velocity, with $b_0(p)$ and $b_2(p)$ defined
by equation (\ref{ecool}) in terms of the upstream and downstream
effective magnetic fields, respectively \citep[\eg][]{webb84}.
This leads to the (test particle) result,
\begin{eqnarray}
p_{ec} = \left(\frac{m_e}{m_p}\right)^{1/2}\left(\frac{6\pi m_ec^2}{\sigma_TcB_{eff,0}^2}\right)^{1/2}\frac{(1-\mathcal{M}_A^{-1})u_s}{\sqrt{\kappa_n q_0\left[1+\frac{1}{\chi}\frac{B_{eff,2}^2}{B_{eff,0}^2}\frac{q_0}{q_0-3}\right]}},\\
 \approx \left(\frac{m_e}{m_p}\right)^{1/2}\left(\frac{6\pi m_ec^2}{\sigma_TcB_0^2}\right)^{1/2}\frac{(1-\mathcal{M}_A^{-1})u_s}{\sqrt{\kappa_n q_0\left[1+\chi\frac{q_0}{q_0-3}\right]}},\nonumber
\label{pec}
\end{eqnarray}
where $u_s(t)=u_0$ is the instantaneous shock speed.
Here we assume $\kappa_0/\kappa_2 = B_2/B_0 = \chi$ and the
second expression corresponds to the limit, $B_{eff} \approx B$,
which is applicable in our simulations.
This $p_{ec}$ translates into a cutoff energy for CR electrons
at the shock,
\begin{equation}
E_{ec} \approx 3.3 ~\rm{PeV}\frac{1}{\sqrt{B_{0,\mu}q_0(1+\chi\frac{q_0}{q_0-3})}}\left(\frac{u_s(t)}{10^4\kms}\right),
\label{eec}
\end{equation}
where the  subscript $\mu$ on B indicates field strength in microGauss.

Equation (\ref{eec}) evaluated with $B_{0,\mu}=30$ and $\chi = 4$ predicts
$E_{ec} \sim 100$ TeV for these models at time $t \sim t_o$, which is consistent
with the simulation results shown in Fig. 4.
This is also consistent with the upper limits on electron spectra derived from observation
for a number of observed SNRs \citep{reyn99} and with SNR simulation results, 
for example, of \citet{bkv02}. 
During the early ST stage, the maximum acceleration momentum 
gradually saturates as given in equation (\ref{pmax}),
but then it eventually asymptotes to $p_{ec}$ determined by the
instantaneous shock speed as given in equation (\ref{pec}). 
As a result, for larger values of $B_0$, the electron cutoff energy is smaller,
but the maximum proton energy is larger. 

Electrons advected into the SNR interior will continue to lose radiative energy with  
the combined synchrotron/IC radiative cooling time, $t_r(p)=p/b(p)$, given by
\begin{equation}
t_r(p)=133~\rm{yrs}\left(\frac{B_{eff}}{100\muG}\right)^{-2}\left(\frac{p_{ec}}{10^4}\right)^{-1}.
\label{trad}
\end{equation}
Setting  the cooling time behind the shock to the SNR age, $t_r(p_{e,1}) = t$,
provides a conservative, rough estimate for the minimum electron momentum influenced by
radiative cooling,
\begin{equation}
p_{eb} \sim 1.3\times 10^3 \left(\frac{B_{eff}}{100~\muG}\right)^{-2}\left(\frac{10^3~yrs}{t}\right).
\label{peb}
\end{equation}
Recall that $E_e \approx p~\rm{GeV}$. Below $p_{eb}$ the integrated electron
spectrum should be very similar in form to the proton spectrum at the same momentum,
consistent with results shown in Fig. 3 and Fig. 4.

To help interpret the integrated electron spectra, $N_e(E)$, between the break momentum $p_{eb}$ and the cutoff momentum
$p_{ec}$, we note in the thin shell
approximation that $N_e(E) \propto n_e(E)r^2\Delta r$,
and that the local electron spectrum, $n_e(p) = K_{e/p} n_p(p)$,
below $p_{eb}$.
In this case it is easy to show that the integrated electron energy spectrum below $E_{ec}$ would
be steeper by about one in the spectral index than the integrated proton spectrum
in the same energy range. 
The steepened slope would extend down to the energy $E_{eb} \approx p_{eb}$ GeV.
On the other hand, in spherical flows
following roughly the ST behavior, the postshock flow accelerates away from the
high magnetic field region just inside the shock while reducing the
magnetic field strength. These effects reduce the rate of
cooling from that implied in equation \ref{peb}, thus shortening the
interval between $p_{ec}$ and $p_{eb}$.
Then the spectral
steepening of the integrated electron population compared to the proton
spectrum will be greater than unity, as observed in our results.

\section{Modeling of continuum emissions from the SNR models}

\subsection{The Cosmicp code for radiative processes}

The nonthermal radio to $\gamma$-ray emissions expected from CR 
electrons and from proton secondary products in the simulated SNRs were computed 
through the post processing of model data using {\it Cosmicp}, a general 
purpose nonthermal continuum emissions code developed in-house around
published radiative process and inelastic particle scattering formulations.  {\it Cosmicp}
computes direct emissions from input electron/positron
populations and calculates the inelastic collision products 
for photon and proton interactions with
matter, including pion and lepton secondaries. 
Lepton secondaries are incorporated into {\it Cosmicp}, 
but they do not make significant contributions to SNR 
emissions, so are not included in our analysis here. {\it Cosmicp} was
designed to be very flexible, so makes as few assumptions as possible.
For example, the input energy distributions of energetic particles
are arbitrary.  Using inputs $n(E,r)$ for CR
electrons and protons along with the relevant environmental information,
such as magnetic field, radiation field,  and gas density,
{\it Cosmicp} calculates the volume emissivity per unit frequency, $j_{\nu}(\nu,r)$, for
each radiative process,
where $\nu$ is the emitted photon frequency.
Bremsstrahlung from the thermal electron population is also included
in our calculations with {\it Cosmicp}, but electronic emissions
involving discrete atomic transitions are excluded. 
All calculations are in cgs units, which allows
easy linking to actual observables.

Specific CR electronic emissions incorporated into our analysis here
include synchrotron, inverse Compton 
(IC), and bremsstrahlung processes.  The synchrotron, IC, and relativistic bremsstrahlung
formalisms followed presentations in \citet{sch02}.  The nonrelativistic 
bremsstrahlung formalism used to compute the spectrum due to the thermal population followed \citet{jackson99}. 
The IC formalism uses an arbitrary, user supplied spectral form for the ambient
photon field, so 
{\it Cosmicp} can model interactions with multiple blackbody or non-blackbody
sources.  It does, however, currently assume an isotropic incident
radiation field.

Hadronic interactions include, as noted, inelastic proton-proton and photon-proton 
interactions.  Photopion production in the latter case can be calculated 
from an arbitrary photon field
using the formalism in \citet{kelner08}.
Pion generation from proton-proton interactions uses relations presented in
\citet{kelner06}.
The principal radiative process that results from hadronic interactions 
in our SNR context is secondary $\pi^0$ decay, which is
dominated by the $2\gamma$ channel.
{\it Cosmicp} can include helium in the hadron-hadron production of
photons. However, that correction was not applied in the present calculations.
For a CR composition similar to that incident on the earth and an ISM
with normal metalicity this
correction would increase the pion production rate by $\sim 25$\%
\citep{sch02}.

\subsection{Nonthermal emissions: Local interaction conditions}
Rates for nonthermal electron bremsstrahlung and p-p collisions leading to pion decay $\gamma$'s
depend on the product of the local plasma density and the CR electron and proton
density, respectively. 
These are taken in our emission calculations
directly from the simulations.
Synchrotron and IC emissions depend on the
spatial magnetic field distribution and the ambient radiation field
respectively.
As discussed in \S 2.2, we assume the local magnetic field strength scales 
with thermal plasma density; i.e.,
$B(r,t)= B_0 [\rho(r,t)/\rho_0]$.
The ambient photon field for our calculations is uniform
and models the
Galactic radiation field in the solar neighborhood as given by 
\citet{sch02}. It consists of a sum of four components: the
Cosmic Microwave Background
(CMB) blackbody at $2.7$ K with an energy density of $0.25 \rm ~eV/cm^{3}$, dust at $20$ K with an energy density of $0.4 \rm ~eV/cm^{3}$, old yellow stars at $5000$ K with an energy density of $0.3 \rm ~eV/cm^{3}$, and
young blue stars at $20,000$ K with an energy density of $0.09 \rm ~eV/cm^{3}$.  The total
energy density for the ambient photon field is $1.04 \rm ~eV/cm^{3}$. In
evaluating equation (\ref{trad}) for electron energy losses this
translates to an effective magnetic field of $B_r = 6.5 \rm ~\mu G$.
We point out in \S3.3.2, however, that IC scattering by the highest energy
electrons of all but the CMB and dust emission radiation fields is
limited by electron recoil, and reduced by the Klein-Nishina cutoff, so this value overestimates
the effective $B_r$. In our models electron losses are dominated by
the synchrotron process, so this is a relatively small correction, in any case.

\subsection{Volume integrated radiation spectrum from the model remnants}

\begin{figure}
   \centering
   \includegraphics[width=0.90\textwidth]{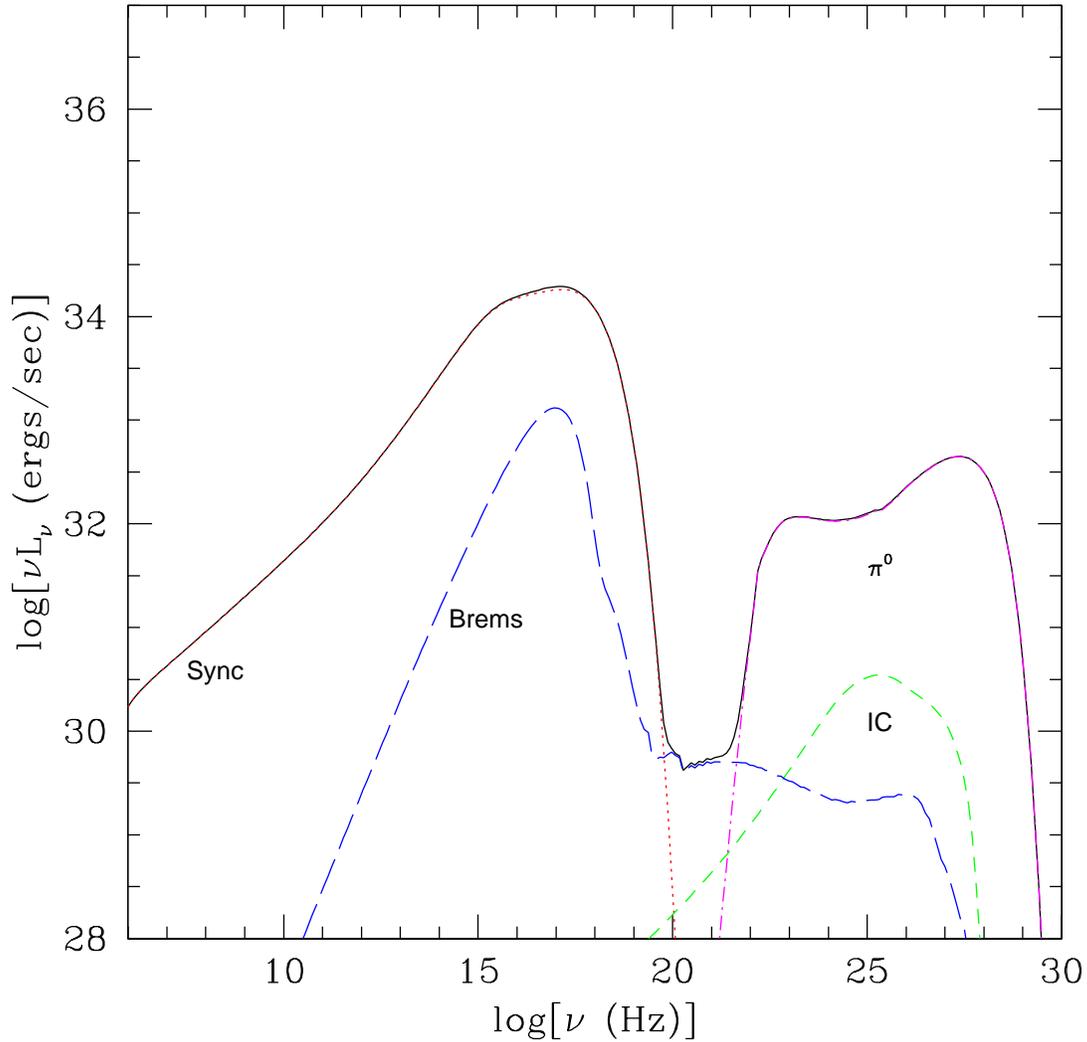}
  \caption{
The volume integrated SED for the S1 model at $t/t_o = 3$: total emission (solid black line),
synchrotron emission (red dotted line), IC emission (green
short dashed line) bremsstrahlung (blue long dashed line),
$\gamma$s from $\pi^0$ decays (magenta dot-dashed line). (A colour version of this figure is available in the online journal.)
}
\end{figure}

A general impression of the radiative emissions produced in the simulated
SN Ia remnants can be obtained from Fig. 5, which shows for
the S1 simulation at $t/t_0 = 3$ the volume integrated
Spectral Energy Distribution (SED), $\nu L_{\nu}$
from radio frequencies to PeV $\gamma$-rays.
Synchrotron emission dominates the spectrum for $\nu \la 10^{20}$Hz\footnote{Useful translation factors are: 1 keV $\rightarrow 2.42\times 10^{17}$Hz,
$10^{17}{\rm Hz~}\rightarrow$ 0.41 keV.}.
There is a narrow $\gamma$-ray window, $10^{20}{\rm Hz}\la \nu \la 10^{21}{\rm Hz}$ dominated
by bremsstrahlung from CR electrons interacting with thermal plasma.
Above $\sim 10^{21}{\rm Hz}$ ($E_{\gamma} \approx 400$keV) the spectrum is dominated by photons from $\pi^0$ decays
for the ISM density assumed in the simulation.

\begin{figure}
   \centering
   \includegraphics[width=0.90\textwidth]{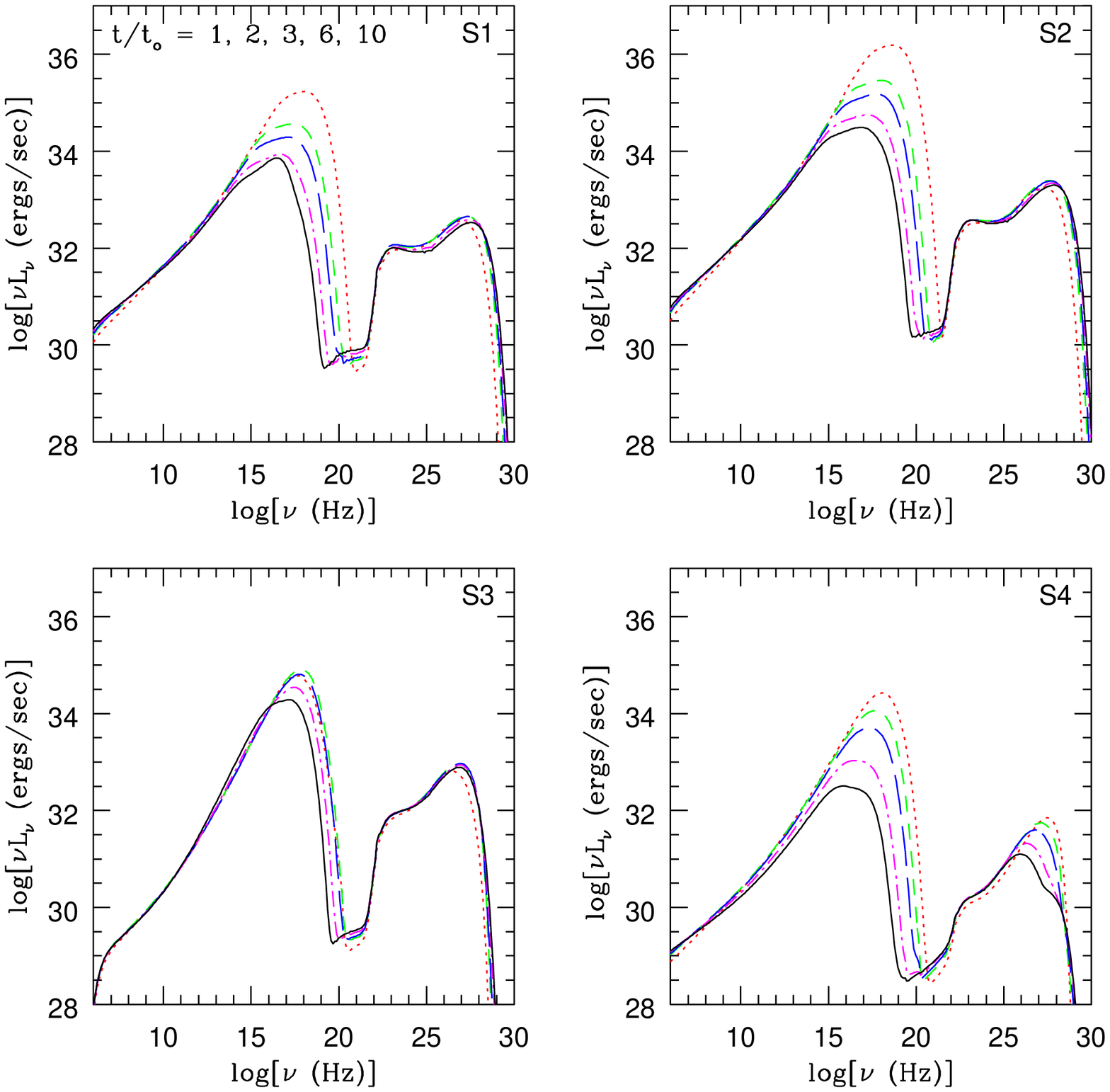}
  \caption{
The total volume integrated SEDs for models S1-S4 over time.  The dotted
red line is $t/t_o = 1$, the short dashed green line is $2$, the long dashed blue
line is $3$, the dot-dashed magenta line is $6$, and the solid black line is
$10$. (A colour version of this figure is available in the online journal.)
}
\end{figure}

The spatially integrated radiative emissions at selected times for all four models 
are shown in Fig. 6. 
Independent of time, the radiation is mostly dominated by
the electronic CR emissions at lower energies, but, for the most part, proton $\pi^0$ decay at high energies. The additional bump in the UV and X-ray seen at late times especially in model S1 is due to thermal bremsstrahlung. 
This feature
appears as the high energy end of the electron distribution is depleted by
radiative losses, revealing thermal emission from hot, postshock plasma. The lower
gas density in the S4 model reduces this thermal contribution, and also
the nonthermal bremsstrahlung $\gamma$-rays.

In the S4  model the $\gamma$-ray portion of the spectrum is a composite, dominated by $\pi^0$ decays between 
$\nu \sim 10^{22} - 10^{23}$ Hz ($E_{\gamma}\sim 40$ MeV - $400$ MeV) and again above about $\nu \sim 10^{28}$ Hz ($E_{\gamma}\sim 40$ TeV). 
However, the lower ambient gas density of the S4 model also reduces the $\pi^0$ production enough that IC emissions
are dominant between the synchrotron cutoff and $\sim 10^{22}{\rm Hz}$.
IC emission is also predominant between $\nu \sim 10^{24} - 10^{27}$ Hz ($E_{\gamma}\sim 4$ GeV - $4$ TeV).

We outline below some useful relations for interpretation of the calculated radiative emissions in the context of
these models, including some simple scaling relations that can be helpful in extending our results to different
model parameter choices than we have used. We begin with discussions of the electronic emissions
and follow with the essential elements to understand the calculated emissions produced by inelastic
p-p collisions.

\subsubsection{Electron Synchrotron Emission}

The electron synchrotron emission spans a frequency range from radio
to roughly the emission peak for electrons near the Lorentz
factor $\gamma_{ec} \approx (m_p/m_e)p_{ec}$. Above this energy
the electron population is heavily modified by losses and begins to drop off severely.
The associated synchrotron peak frequency is 
$\nu_{sc} \approx 0.3~\nu_{crit}$, where 

\begin{equation}
\nu_{crit} = \frac{3}{2} \nu_B \gamma_{ec}^2 \sin (\theta)
= 4.2 \times 10^{16}~ {\rm Hz} \left( \frac{B}{100 \mu{\rm G}}\right)
 \left( \frac{\gamma_{ec}}{10^7}\right )^2 \sin (\theta),
\label{nusync}
\end{equation}
and $\nu_B$ is the nonrelativistic electron cyclotron frequency \citep{bg70}.
For the calculations presented here 
we set the angle of the magnetic field with respect to the observer,
$\theta = 90^o$. 

Using equations (\ref{pec}) for the upper electron energy cutoff and (\ref{nusync}) we can derive the following estimate for the cutoff frequency in the synchrotron
spectrum in the limit that $B_{eff} \approx B$; namely, 

\begin{equation}
\nu_{sc} \approx 1.8 \times 10^{20}~ {\rm Hz}~ 
\frac{1}{q_0(1+\chi\frac{q_0}{q_0-3})}\left(\frac{u_s(t)}{10^4\kms}\right)^2.
\label{nusyncpec}
\end{equation}
For $\chi = 4$ and $q_0 = 4$ the factor $q_0[1+\chi q_0/(q_0-3)] = 17$,
so $\nu_{sc} \approx 3\times 10^{18}~\rm{Hz}~(u_s/10^4 \rm{km/s})^2$,
consistent with the results in Fig. 6.
As noted by previous authors \citep[\eg][]{bamba03,berez04}
the cutoff frequency is determined mainly the instantaneous shock
speed and independent of magnetic field strength except through 
its implicit dependence on the factor $\chi$. 
This explains the similarity between the values for the synchrotron cutoffs for S1 and S3 even though they have very different magnetic fields. 

The break frequency that corresponds to the estimated break momentum $p_{eb}$ 
depends only on magnetic field strength and the shock age as
\begin{equation}
\nu_{sb} \approx 7.52\times 10^{14} {\rm Hz}
 \left(\frac{10^3~\rm{yrs}}{t} \right)^2 \left(\frac{B_{\rm eff}}{100\muG}\right)^{-3}
\label{nusb}
\end{equation}

The values of $\nu_{sc}$ and $\nu_{sb}$ generated from equation (\ref{nusyncpec})
and (\ref{nusb}), respectively,
are consistent with the SEDs shown in Figs. 5 and 6.
The photon energies corresponding to the synchrotron cutoff frequency 
fall in the range $0.1-10$ keV, consistent with observed Type Ia remnants \citep{reyn99}.

The synchrotron luminosities of all four simulations below the X-ray cutoffs can be approximately related to each other 
by a simple scaling relation that, in addition, provides a means to extend
these results approximately to different parameter choices.  
For $E<10^2 {~\rm GeV}$ the volume integrated electron energy spectrum shown in Fig. 4 can each be represented by
\begin{equation}
N_e(E,t) \approx (s-1)\mathbf{N}_{e}E_i^{s-1} E^{-s},
\label{edistr}
\end{equation}
where $E_i \sim k_B T_2$ represents the suprathermal injection energy of CR electrons, $\mathbf{N}_{e}$ is the volume integrated number of CR electrons
and $s$ is the energy distribution mean power law slope between $E_i$ and $<E<10^2{~\rm Gev}$.
For reference, recall that momentum and energy distribution
power law indices are related as, $s = q-2$.
The synchrotron luminosity can then be expressed approximately as  \citep{bg70}
\begin{equation}
\nu L_{\nu, sync} \propto \mathbf{N}_e~E_i^{s-1}~ B^{(s+1)/2} \nu^{-(s-3)/2}.
\label{synchp}
\end{equation}
Since $\mathbf{N}_e  \propto \xi K_{e/p} M_{sw}$ 
and the injection faction $\xi$ is nearly a constant  
for $t/t_o \ga 1$ in these simulations,
$\mathbf{N}_e \propto M_{sw} \propto (t/t_o)^{6/5}$.
The injection energy, $E_i$ scales roughly with the shock speed, so also as
$E_i \propto M_{sw}^{-1/2}$.
In \S 2.2, we established the approximate scaling, $\chi \propto M_{sw}^{-1/6}$,
so $B\propto M_{sw}^{-1/6}$.
Then $\nu L_{\nu, sync} \propto M_{sw}^{(17-7s)/12}$ in these simulations.
For example, $\nu L_{\nu,sync}\propto M_{sw}^{1/4}$ for $s = 2$ or $\nu L_{\nu,sync}\propto M_{sw}^{1/18}$ for $s = 7/3$. 
Since the index $s$ varies slowly within the range $9/5 \la s \la 7/3$
in these simulations, the radio luminosity should depend very weakly 
on $M_{sw}$ or $t/t_o$, consistent with the results in Fig. 6.

Note that equation (\ref{synchp}) alone is not sufficient to estimate synchrotron emission behaviors within the X-ray band.
The synchrotron emission in this band is generated predominantly by electrons near the spectral
break $\nu_{sb}$ and cutoff $\nu_{sc}$. 
Consequently, X-ray synchrotron emissions are heavily influenced by 
the positions of these frequencies. 

\subsubsection{Electron Bremsstrahlung and IC Emission}

The bremsstrahlung and IC contributions are mostly subdominant in our models.
At early times synchrotron emission dominates thermal bremsstrahlung over
the bands where they both contribute.
The bremsstrahlung peak is roughly $\nu_{T} \sim kT_2/h$, while
the synchrotron cutoff, $\nu_{sc}$, is given by equation (\ref{nusyncpec}). 
Both cutoffs depend primarily on the shock speed squared, so 
evolve more or less together. 
The decreasing role of synchrotron emission in the X-ray band results,
on the other hand, from the fact that CR electron energy losses downstream of the shock
cause that population to become further depleted, dropping the relative emitted power
between $\nu_{sc}$ and $\nu_{sb}$,
revealing the thermal bremsstrahlung emission.  Consequently for $t/t_o \ga 6$ 
in S1 thermal
bremsstrahlung is an important X-ray contributor (see Fig. 6).

In another exception to the general emission behaviors, we note that IC emissions near $h\nu \sim1$ TeV
in model S4 exceed 
the otherwise dominant pion decay gamma-rays. 
The S4 $\pi_0$ decay emissions, on the other hand,
are smaller by two orders of magnitude, reflecting the similarly lower S4 ambient density 
(see equation [\ref{pip}]).

Some comments on the contributions and form of the IC SED may be useful.
As noted in \S 3.2 the IC emissions result from a combination of incident photon fields
with blackbody forms at different characteristic temperatures. 
The effective incident field is just their sum. 
While the CMB is a blackbody in both
spectral form and energy density, the radiation from dust and stars has
greatly diluted energy density compared to the Planck functions for their respective temperatures. 
In each case we can assign a frequency-independent
dilution factor, $D$,
given by the ratio of the local energy density to the appropriate Planck function. The dilution factors
for the radiation field properties listed in \S 3.2 are: for the CMB, $D_{CMB} = 1$, for
thermal dust emission, $D_d = 5.3 \times 10^{-4}$, yellow stars, $D_{y*}=1.0 \times 10^{-13}$,
and for blue stars, $D_{b*}=1.2 \times 10^{-16}$.
 
The IC scattering source function in the limit of Thomson scattering 
of incident photons with a Planck spectrum with color temperature, $T_i$, and
dilution factor, $D_i$, is 
\begin{equation}
q_{IC,i}(E_{\gamma} \gg k_B T_i \gamma_{ec}^2 ) = \frac{r_0^2 (s-1)~p_i^{s-1} }{\pi \hbar^3 c^2} K_{e/p}~\mathbf{n}_p~
F(s) \left(\frac{m_p}{m_e}\right)^{s-1} (k_B T_i)^{(s+5)/2} D_i E_{\gamma}^{-(s+1)/2}  
\label{qic}
\end{equation}
where $r_0$ is the classical electron radius,
$F(s)$ is a dimensionless scaling function as defined in \citet{bg70}; 
for $s\sim 2$ $F(s) \sim 5$.
We
have used the form of the electron spectrum given in equation (\ref{edistr}).

For a representative CR electron slope, $s = 2$, the different background radiation
components contribute in equation (\ref{qic}) according to factors $\propto T_i^{7/2}D_i$, which
leads to a ranked list $CMB,d,y*,b*$ represented in the ratios $1:0.58:0.03:0.004$.
This suggests that the CMB and the dust IR radiation are the predominant contributors to the IC 
SED, despite the roughly comparable photon energy densities of the stellar components (see \S 3.2). 
In this regard it is also worth noticing that the IC spectral peak at a few TeV involves
photons scattered to energies close to $E_{ec}$, the maximum possible, whereas the
Thomson scattering limit $E_{\gamma}\sim k_B T_i\gamma_{ec}^2$ would take photons into
the PeV range when scattered from the incident stellar photon fields. In fact, photons from those
higher temperature incident fields are scattered in the Klein-Nishina limit by the most
energetic electrons, as given by the condition on
the incident photon energy, $h\nu >m_ec^2/\gamma_{ec}$. This condition can be expressed
practically as
\begin{equation}
T_i > \frac{(m_ec^2)^2}{k_B E_{ec}}\approx 3.0\times 10^3\left(\frac{\rm TeV}{E_{ec}}\rm \right) ~\rm Kelvin.
\label{kn}
\end{equation}
For $E_{ec}\sim 100~{\rm TeV}$ only the CMB and, marginally, the
dust radiation escape this limit. The scattering cross section in the Klein-Nishina 
limit is reduced from the Thomson cross section, so that, in fact,
gamma ray IC emissions in these models are dominated by the CMB and dust emissions.

We can use equation (\ref{qic}) to construct an approximate, simple scaling relation for the IC luminosity, $\nu L_{\nu, IC}$, in the Thomson regime assuming an electron
distribution of the form in equation (\ref{edistr}),
\begin{equation}
\nu L_{\nu, IC} \propto \mathbf{N}_e E_i^{s-1} u_{rad}^{(s+5)/8} \nu^{-(s-3)/2}
\label{icp}
\end{equation}
where the radiation energy density, $u_{rad}$, would be approximated by the sum of the CMB and dust radiation
densities. 
The IC luminosity, $\nu L_{\nu, IC}$, scales as $M_{sw}^{(3-s)/2}$, similar to the synchrotron luminosity . 
In fact, the ratio of the two for emissions at synchrotron frequency $\nu_{sync} \ll \gamma_{ec}^2\nu_B$,
and IC frequency, $\nu_{IC} \ll \gamma_{ec}^2\nu_{T_d}$ in, say, the
radio and gamma ray bands, is just
\begin{equation}
\frac{\nu L_{\nu, IC}}{\nu L_{\nu, sync}} \propto \frac{u_{rad}^{(s+5)/2}}{u_B^{(s+1)/4}} 
\left(\frac{\nu_{sync}}{\nu_{IC}}\right)^{(s-3)/2}
\label{icsync}
\end{equation}
This indicates that the ratio between the two contributions to the SED for each
of our SNR models remains roughly constant over time. Since $u_{rad}$ is
model independent, the ratio of IC gamma ray luminosity to radio synchrotron
luminosity varies in these models roughly  as $\propto 1/B_o^{3/2}$.

\subsubsection{$\pi^0$ Decay Gamma-Ray Radiation}

Gamma ray decay products from inelastic p-p collisions, and $\pi_0 \rightarrow 2\gamma$,
in particular, are commonly viewed as the ``smoking gun'' for hadronic DSA in SNRs.
As noted above these emissions mostly dominate the gamma ray SEDs in the SNR
models being discussed here.
Although the detailed modeling of $\pi^0$ decay gamma rays is fairly
complicated, a simple analytic approximation offered by 
\citet{pfr03} provides a good understanding of the $\pi^0$ gamma ray SED
in these models. In particular, if the local CR proton energy spectrum 
above the threshold for pion production ($E>780\rm MeV$) is
a power law, $n_p(E) = (s-1)\mathbf{n}_pE_i^{s-1}E^{-s}$, then the $\gamma$-ray source function,
$q_{\pi}$, (photons/time/volume/energy) is
\begin{equation}
q_{\pi} \approx \frac{\sigma_{pp} c}{m_pc^2} 2^{2-s} \frac{4}{3s} \left ( \frac{m_{\pi}}{m_p}
\right )^{-s} \left [ \left ( \frac{2E_{\gamma}}{m_{\pi}c^2} \right )^{\delta} 
+ \left ( \frac{2E_{\gamma}}{m_{\pi}c^2} \right )^{-\delta} \right ]^{-s/\delta} 
~n~(s-1)E_i^{s-1}\mathbf{n}_p 
\label{qpi}
\end{equation}
where $ \sigma_{pp} \approx 3.2 \times (0.96+e^{4.4-2.4s}) \times 10^{-26} \rm ~cm^2$  
is an effective p-p cross-section for pion production, 
$\delta = 0.14s^{-1.6}+0.44$,  
while $n$ is the density of the thermal protons, and
$m_{\pi} = 135 \rm MeV/c^2$ is the $\pi^0$ mass. 

This source function peaks for $E_{\gamma} = m_{\pi}c^2/2 = 67.5~$MeV.
At low energies it asymptotes to a power law $q_{\pi} \propto E_{\gamma}^s$,
while it asymptotes at high energies to a power-law $q_{\pi}\propto E_{\gamma}^{-s}$.
The source function cuts off above roughly 10\% the proton high energy
cutoff (Kelner et al. 2006). 
The radiated power per unit volume (the SED) will scale with $E_{\gamma}^2~q_{\pi}$. Using
equation (\ref{qpi}) for $s>2$ the SED would peak around $E_{\gamma}=67.5~\rm{MeV}~[(\delta+2\delta/s)/(\delta-2\delta/s)]^{1/(2\delta)}$.
There is no SED maximum for $s\le 2$, except at the high energy cutoff.

Assuming a proton spectrum  of the form in equation (\ref{edistr}) one obtains
a simple high energy scaling relation for the volume integrated pion decay luminosity
\begin{equation}
\nu L_{\nu, \pi^0} \propto n \mathbf{N}_p ~ E_i^{s-1}~E_{\gamma}^{2-s} \propto \xi M_{sw} E_i^{s-1}~ n ~E_{\gamma}^{2-s}. 
\label{pip}
\end{equation}
This scaling is consistent with the results shown in Fig. 6, showing in particular
why the pion decay emissions are similar in all the models except for S4 because of
its lower $n_{ISM}$.

The proton distributions in these SNR models are not really true power laws, 
of course, but
have concave upwards forms; that is, $s$ decreases with energy. Protons near the pion production
thresholds just above a GeV are mostly responsible for gamma rays near the 
low energy gamma ray peak. In that range
typically $s \sim 2.3$ (see Figs. 3 and 4), The expected
$\pi^0$ gamma ray SED peak is near 1 GeV ($\nu \sim 2\times 10^{23}$Hz), consistent with
Figs. 5 and 6.
The model $\pi^0$ SEDs in Figs. 5 and 6 also exhibit a second, higher energy
peak. That results from the concave upwards character of the proton spectra
and their eventual cutoffs around or above a PeV for $t/t_o>1$. Except for
model S4 the proton spectra have $s<2$ below their cutoffs, making their
upper $\pi^0$ gamma ray peaks quite prominent.

Since the the upper $\pi^0$ gamma ray peak is so prominent it can be used to determine the peak proton energy in the SNR \citep[\eg][]{aha07}.  From the gamma-ray production relations in Kelner et al. (2006) one can see that the peak in gamma-ray production for a monoenergetic proton of $E_p> 1 {\rm TeV}$ occurs at a photon energy $E_{\gamma}\approx 0.1E_p$. Looking at Fig. 6 for S1-S3 the gamma-ray peak energy is around about $40-400$ TeV ($10^{28}-10^{29}$ Hz) which corresponds to a proton energy of $400-4000$ TeV.  This is consistent with the peaks in the proton spectra in Fig. 3.

Both IC and $\pi^0$ source functions
depend on the number of CRs accelerated, while the pion-generated
gamma rays also depend on the density of the thermal plasma. 
The ratio of $q_{\pi}/q_{IC}$
found from equations (\ref{qpi}) and (\ref{qic}),
\begin{eqnarray}
\frac{q_{\pi}}{q_{IC}} = \pi \frac{\sigma_{pp}}{r_0^2}\frac{\hbar^3 c^3}{(k_BT)^2m_ec^2}
 \frac{4\cdot2^{2-s}}{3s}\left (
\frac{m_{e}}{m_{\pi}} \right )^s \left(\frac{E_{\gamma}}{k_BT}\right)^{(s+1)/2} \nonumber \\
 \cdot \left [ \left
( \frac{2E_{\gamma}}{m_{\pi}c^2} \right )^{\delta} + \left (
\frac{2E_{\gamma}}{m_{\pi}c^2} \right )^{-\delta} \right ]^{-s/\delta}
\frac{n}{D F(s) K_{e/p}},
\label{ratpiic}
\end{eqnarray}
provides a useful comparison of the two processes for estimation 
and scaling purposes. 

Again, taking $s = 2$ as representative, then including CMB and dust contributions
in the radiation field as discussed earlier, and setting $K_{e/p} = 10^{-4}$
along with $n=2 ~\rm cm^{-3}$, appropriate
to postshock conditions for model S1 (see Table 1 and Fig. 2), we can use equation (\ref{ratpiic})
to estimate 
\begin{equation}
\frac{q_{\pi}}{q_{IC}} \sim 400  \left(\frac{E_{\gamma}}{400{\rm MeV}}\right)^{-1/2} 
\left(\frac{n}{2{\rm cm^{-3}}}\right) \left(\frac{10^{-4}}{K_{e/p}}\right),
\label{pic}
\end{equation}
which is valid roughly for $E_{\gamma}\sim 400$MeV ($\nu \sim 10^{23}$Hz).
This number is in reasonable agreement with the numerical results displayed
in Fig. 5. 
Equation (\ref{pic})
suggests that $\pi^0$ decay emissions would need to be reduced by at
least two orders of magnitude from those
in model S1 before IC emissions would become dominant in the GeV band.
Note for scaling purposes from equation (\ref{ratpiic}) that
$q_{\pi}/q_{IC} \propto n/K_{e,p}$. Consequently, GeV IC emissions
would dominate when $n_{ISM}\la 30 K_{e,p}~\rm{cm}^{-3}$, consistent
with the model S4 results as displayed in Fig. 6.

\subsection{X-ray and gamma-ray brightness distributions}

\begin{figure}
   \centering
   \includegraphics[width=0.95\textwidth]{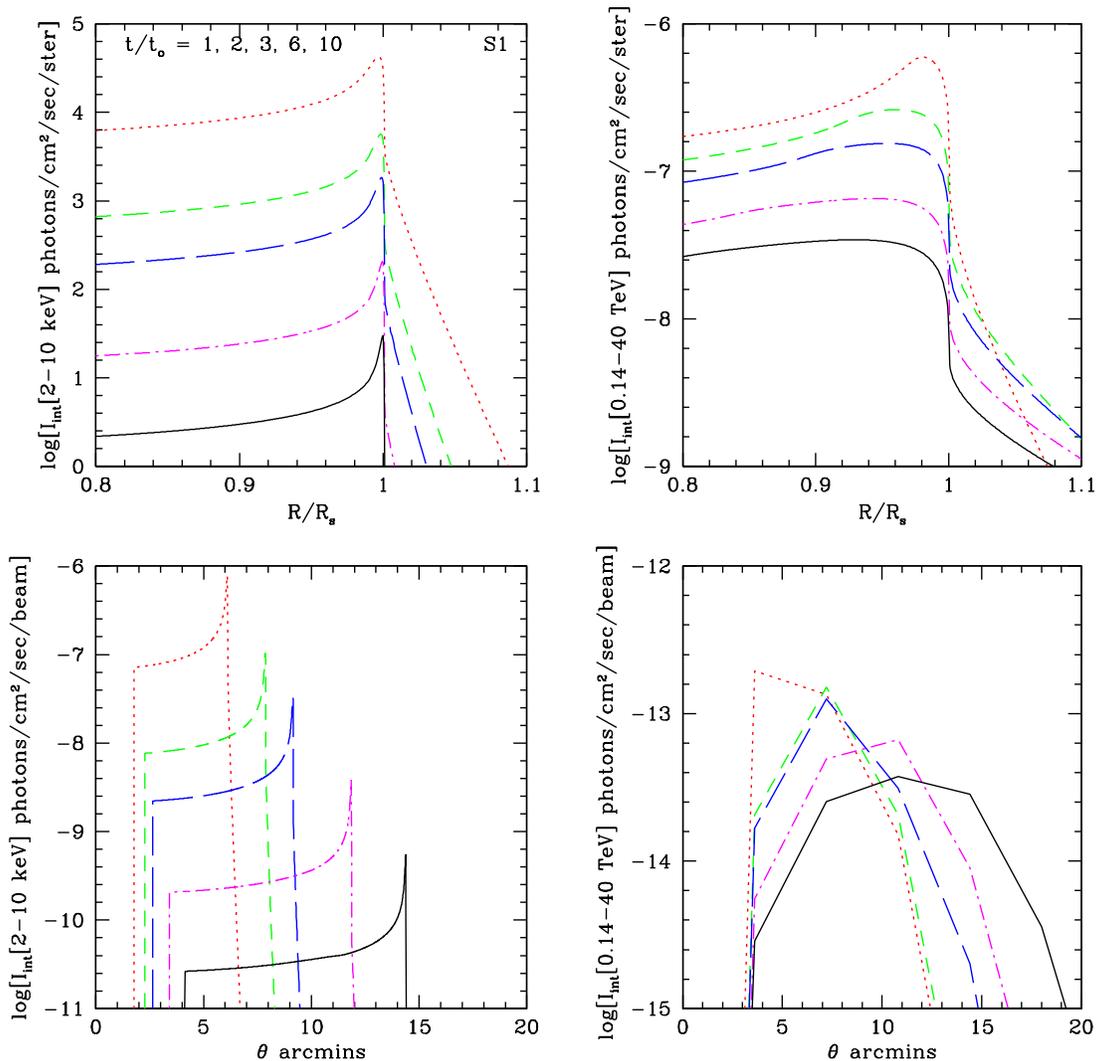}
  \caption{
The radial brightness profiles for the S1 model in the {\it Chandra} X-ray band ($2$ - $10$
keV) and HESS band ($140$ GeV - $40$ TeV). Upper panels: Brightness profiles in terms
of projected radius, $R/R_s$. Lower panels: The same profiles in angular
units convolved
to the resolution of {\it Chandra} and HESS with the SNR at $2$ kpc. The line types are the same as Fig. 5. (A colour version of this figure is available in the online journal.)
 }
\end{figure}

Since X-ray telescopes and the new generations of gamma-ray telescopes have the 
ability to
resolve images of SNRs with distances inside some tens of kpc, it is valuable to model spatial 
brightness profiles of these emissions.  The profiles also provide insights to the
similarities and differences among the different emission processes.
The spatial profiles for all 4 models look very similar, so we will focus on model S1.
The top panels of Fig. 7 show at several times the model S1 radial brightness profiles of broadband fluxes in the 
{\it Chandra} X-ray band ($2$ - $10$ keV) \citep{garm03} and the High Energy Stereoscopic System (HESS)
gamma-ray band ($140$ GeV - $40$ TeV) \citep{ber07}.  
The profiles are plotted against the radius perpendicular to the line of sight from the center of the SNR, what we will call the projected radius $R$,
relative to the projected shock radius, $R_s(t)$.
The lower panels of Fig. 7 present the same profiles in angular units
when the SNR is placed at a distance of 2 kpc and the image has
been convolved to the nominal resolutions in these bands for Chandra and HESS.

Since the exponential cutoff of the synchrotron radiation spectrum, 
$E_{\gamma}(\nu_{sc})\propto (t/t_o)^{-6/5}$ lies in  the {\it Chandra} band, 
the $2$ - $10$ keV X-ray fluxes decrease quickly in time.
This is evident in the evolution of the X-ray brightness profiles in Fig. 7.

The X-ray brightness profiles are also very narrowly peaked at the projected shock position.
The X-ray profiles in Fig. 7 have characteristic apparent FWHM (Full Width Half Max) of $\sim 1$\%.
This is consistent with X-ray observations of several young SNRs that have revealed thin nonthermal filaments
at the forward shock with a FWHM thickness $\sim 1\%$ of the shock radius 
in projection \citep{bamba03, ballet06,pari06}.  
One can estimate the observed thickness of the shock rim when radiative losses are the
dominant factor by the following relation from \citet{pari06},
$\Delta R_{obs} = 4.6 u_2 t_r$,
where $t_r$ is defined in equation (\ref{trad}), and the factor $4.6$ is a geometrical correction that comes from projecting 
a spherical shock with an exponential emission profile onto a plane and taking the FWHM of the resulting profile \citep{ballet06}.
Taking $u_2= U_{\rm ST}/\chi$, the ratio of shock speed to
shock radius for the ST similarity solution, $U_{\rm ST}/r_{\rm ST} = (2/5)/t$,
the relative FWHM shock thickness can be approximated as 

\begin{equation}
\frac{\Delta R_{obs}}{R_s} 
\approx \frac{1.84}{\chi} \left( \frac{t_r}{t} \right) 
= \frac{0.11}{\chi}  \left( \frac{B_{\rm eff}}{200 \mu{\rm G}}\right)^{-2}
 \left( \frac{\gamma_{ec}}{2\times10^7} \right ) ^{-1}
 \left( \frac{t}{500 {\rm yrs} }\right ) ^{-1}, 
\label{robsratio}
\end{equation}
where the numerical evaluation of $t_r$ from equation (\ref{trad}) has been applied
on the RHS.
Taking $\chi \sim 7$ for young SNR along with characteristic values for
the rest of parameters, equation (\ref{robsratio}) indeed predicts $\Delta R_{obs}/R_s \sim 1$\%.
We note, in addition, applying equation \ref{pec} in equation \ref{robsratio}
that $\Delta R_{obs} \propto B^{-3/2}$, independent of time or the
SNR shock speed (\citep[see, also, e.g.,][]{vbk05}.

The gamma-ray spatial profile in Fig. 7 is dominated by $\pi^0$ decay and 
is very broadly peaked at the shock.  The peak at the shock is due to the emissivity scaling as the product of the
CR density and the thermal gas density. Roughly, this will resemble $j_{\nu,\pi}\propto \rho^2$, once again similar to
thermal bremsstrahlung.  However, the brightness profile is considerably broader than the emissivity owing to same projection effects noted earlier for the X-ray profile.

The IC spatial profile, which is mostly buried under the neutral pion radiation in
these models, will have a similar profile to the X-ray synchrotron, because 
they both come from the same population of electrons. Thus the same argument for why the synchrotron rim should be thin would apply to the IC spatial profile. However, the X-ray synchrotron profile will be slightly thinner than the IC profile owing to its dependence on the magnetic field. Since the IC rim will be thin and the $\pi^0$ decay rim will be thick one could in theory differentiate between the two scenarios by looking at the shock thickness. If the shock is thinner in the TeV gamma-rays than the thermal gas distribution, like the X-ray synchrotron profile, then IC is dominant. But if the $\gamma$-ray profile has a thickness similar to the gas profile thickness, then $\pi^0$ decays are dominant.  
This level of discrimination, unfortunately, will not be available in
the near future.

The lower panels of Fig. 7 provide a look at the S1 model SNR placed at a
distance of $2$ kpc, about the distance to SN 1006. The intensity profiles have been
convolved, as outlined below, to the angular resolution of 
the {\it Chandra} ACIS (Advanced CCD Imaging Spectrometer) in the X-ray ($0.5$ arcsec \citet{garm03}), 
and HESS around one TeV ($3.6$ arcmin \citet{ber07}). 
The {\it Chandra} ACIS has
a diffuse flux sensitivity ($3 \sigma$ detection) of $1.6 \times 10^{-5} \rm counts/cm^2/sec/arcsec^2$  or about $1.3\times 10^{-5} \rm counts/cm^2/sec/beam$
for $10$ ksecs of
observing time. Our model X-ray rims would be fully resolved by Chandra at the
2 kpc distance, although none of the earliest profile shown would be detected with
a 10 ksec ACIS exposure.

HESS resolution is sufficient to identify a shell-like structure of the
gamma-ray profile, but would not be able to resolve the width of the rim, since
only a few beams would fit across the source. 
The HESS diffuse flux sensitivity  ($8 \sigma$ detection) of about $2.3 \times 10^{-14}
\rm counts/cm^2/sec/arcmin^2$ or about $9.3 \times 10^{-13} \rm counts/cm^2/sec/beam$ for $103$ hours \citep{garm03, bkv09} would
be sufficient to detect this model source at each time shown.
Detailed comparison to models of the gamma-ray rim will require the higher resolution of next generation of Imaging Atmospheric Cherenkov Telescope's (IACT) such as the Cerenkov Telescope Array (CTA) \citep{wagner09}.

\subsection{Comparisons to real SNRs}

We round out this discussion by referencing the emissions from these model
SNRs to some real Type Ia SNRs. In doing so we emphasize that the models
S1-S4 were not intended to reproduce any particular SNR, but rather
to represent properties of CR modified Type Ia SNRs evolving in typical
environments during the ST stage. The following exercise aims not to
establish precise physical properties of the selected real SNRs, but to
illustrate how broadband observations, along with simple scaling relations,
can be used in conjunction with such models for
a qualitative assessment of the physical state of an observed
remnant.  The potential for such applications is growing greatly with
each new generation of high energy observatory. For instance
several SNRs have been observed by IACTs at TeV photon energies, now including the
Type Ia SN 1006 \citep{acero10} and Tycho \citep{acciar11}. Similarly,
{\it Fermi} is collecting an impressive catalog of SNRs detected in  
the $0.2 - 300$ GeV gamma ray band \citep[\eg][]{abdo10}.  {\it Fermi} has not reported
detection of any Type Ia SNRs so far. However, the core collapse SNR, Cas A, which has been seen
at a distance of 3.4 kpc, provides a useful benchmark.  
{\it Fermi} measured a flux,
$\nu F_{\nu} \approx 5 \times 10^{-12} ~{ \rm ergs/cm^2/sec}$ 
at $12.2 \sigma$ significance after 396 days of all sky observation \citep{abdo10}. 
This translates to an isotropic luminosity, $\nu L_{\nu} = 7 \times 10^{33} ~{\rm ergs/s}$.  
Over an equivalent time {\it Fermi} would not detect our remnants at the distance of Cas A.  However {\it Fermi} is scheduled to observe in all sky mode for at least 5 years, so it could marginally detect ($\sim 3 \sigma$) models S1-S3 at $t/t_o \sim 10$ at the distance of Cas A.

As noted in \S 3.3.3, the TeV gamma-ray spectra of these model SNRs
extend to photon energies well above the upper limits for IACT's such as HESS of 
$\sim 40$ TeV.  The proposed High Altitude Water Cherenkov experiment (HAWC) would 
be able to reach above this limit and would have a $5 \sigma$ detection threshold 
for photons above 1 TeV of $\nu F_{\nu} \approx 1 \times 10^{-12} ~{ \rm ergs/cm^2/sec}$ for $1$ year \citep{gon07}.  
At a distance of $3$ kpc that translates to a luminosity of $\nu L_{\nu} = 10^{33} ~{\rm ergs/s}$.  This would allow HAWC to detect all of our models excluding S4 at that distance after $1$ year.

For the other three primary bands, radio, X-ray and TeV $\gamma$-ray, actual observations of Type Ia SNRs already exist.  We use the scaling relations in the 
previous sections to compare our models to two well observed Type Ia SNRs; SN 1006 and Tycho.
We caution again that the simulations were idealized, so we should not expect
close matches. Still, the comparisons may offer simple insights into the dynamical
states of the SNRs and their CR populations.

\subsubsection{SN 1006}
SN 1006 lies at a distance of about $2.2$ kpc 
and at that distance had a radius of $9.6$ pc as of $1993$ (t=$987$ yrs) \citep{bkv09}.
Ambient density estimates vary, but the recent proper-motion-based dynamical
estimate,  $n_{ISM} \approx 0.085~\rm{cm}^{-3}$, \citet{katsu09} is representative. 
In order to establish a dynamical state for the comparison we utilize the ST expressions in \S 2.2 to obtain

\begin{equation}
\xi_s = \left ( \frac{r_s}{r_o} \right ) \left ( \frac{t_o}{t} \right )^{2/5} = \frac{r_s}{t^{2/5}} \left ( \frac{\rho_o}{E_o} \right )^{1/5}.
\label{xiscale}
\end{equation}
 
The age, t, is well constrained,
and the shock radius, $r_s$, is relatively well constrained by observation. 
Thus, even though
$\rho_o$ and $E_o$ are not individually as well constrained, their ratio
$\rho_o/E_o$ is fixed at least within the ST paradigm. Since the
explosion energy is likely close to $10^{51}$erg (one ``FOE''), by assuming
an ejected mass, $M_{ej} = 1.4M_{\odot}$, we can express the
ambient density constraint as $n_{ISM} \approx 0.037~E_{FOE}~\rm{cm}^{-3}$,
consistent with the \citet{katsu09} estimate.
Similarly, $t_o = 520~E_{FOE}^{-5/6}~$yrs, so at the observed time, $t/t_o = 1.9~E_{FOE}^{5/6}$. 
The swept up mass, $M_{sw} \approx 4.6 M_{\odot}E_{FOE}$. We can use our simulation
S1 for comparison with SN 1006.

The model S1 1.4 GHz radio luminosity, $\nu L_{\nu} \approx 3\times 10^{31}E_{FOE}~ {\rm ergs/sec}$.
The measured 1.4 GHz flux of SN1006 is $\nu F_{\nu}\approx 2.2 \times 10^{-13}~ \rm ergs/cm^2/sec$ \citep{reyn92},  
which translates to an isotropic luminosity, $\nu L_{\nu} \approx 1.3 \times 10^{32}~ \rm ergs/sec$, so slightly larger than the model. As noted in \S 3.3.1 the
model radio emission is almost independent of time, or swept up mass, but it
does scale with the electron injection efficiency, $K_{e/p}$
and the magnetic field, $B$, as $K_{e/p} B^{3/2}$ (assuming $s = 2$). So, a modest
increase in this combination would bring the two luminosities into satisfactory
agreement.

The expected X-ray synchrotron luminosity is set within the model by the radio luminosity
and by the break frequency, 
$\nu_{sb} \propto E_o/M_{sw}^{5/3}$. For model S1 at $t/t_o \approx 2$ the break frequency, $\nu_{sb} \sim 10^{17}$Hz, compared
to the ``roll-off'' frequency in SN 1006 estimated by \citet{bamba08} to be $\nu_{roll} \sim 6\times 10^{16}$Hz.
Fig. 6 shows the model S1 X-ray luminosity near the break to be roughly $10^{34}$erg/sec.
The measured $0.1 - 2$ keV (mostly nonthermal) X-ray flux of SN 1006 is 
$\nu F_{\nu} \approx 1.42 \times 10^{-10}~ \rm ergs/cm^2/sec$ \citep{allen01}.  
This translates to an isotropic luminosity, $\nu L_{\nu} \approx 8.1 \times 10^{34}~ \rm ergs/sec$.
The X-ray comparison is consistent with the radio comparison; the best
match would come from a modest increase in  $K_{e/p} B^{3/2}$.

HESS has reported detection of VHE gamma-rays from SN 1006 with a flux
above 1 TeV gamma ray flux, $\nu F_{\nu}\approx 3 \times 10^{-13}~ {\rm ergs/cm^2/sec}$ \citep{acero10}.  
This translates to a luminosity of $\nu L_{\nu} \approx 2 \times 10^{32}~ {\rm ergs/sec}$.  
This is very close to the (pion-decay) luminosity for model S1 shown in Fig. 6.

\subsubsection{Tycho}

Tycho's SNR may not yet have reached the ST dynamical stage \citep{vbk08}, but we
include it in order to illustrate another application of the scaling relations. 
In this context we note that \citet{kang06} found evolution of the
outer, blast wave, shock to be represented reasonably well by ST scaling well
before the actual ST phase begins.

Tycho has a distance that is variously estimated, but mostly less than about $3.6$ kpc.
We will parameterize our comparison including a distance factor $d_{3.6} = d/3.6 kpc$. 
The corresponding radius in $1983$ (t=$411$ yrs) was $r_s= 4~d_{3.6}~$pc \citep{vbk08}.
The SNR lies in the plane of the galaxy in a warm ISM region of
estimated density $\sim 0.3 ~{\rm cm^{-3}}$.  The warm ISM
context makes models S1 - S3 the most appropriate 
comparison. Since we have established the validity of the emission scaling
relations among these models, we focus on model S1.
We follow the same procedure as for SN 1006, except we now include
distance as an adjustable parameter in order to illustrate its
influence on the comparison. Equation (\ref{xiscale})
leads in this case to $n_{ISM} = 0.43 ~E_{FOE}/d_{3.6}^5~\rm cm^{-3}$
along with $t/t_o = 1.8 ~E^{5/6}_{FOE}/d_{3.6}^{5/3}$ and $M_{sw}\approx 4.3 M_{\odot}E_{FOE}/d_{3.6}^2$.

Tycho's measured $1.4$ GHz radio flux is $\nu F_{\nu} \approx 1.6 \times 10^{-13}~ \rm ergs/cm^2/sec$ \citep{reyn92},  
which translates to an isotropic luminosity, 
$\nu L_{\nu} \approx 2.5 \times 10^{32}~d_{3.6}^2~ \rm ergs/sec$. The scaling 
given in equation (\ref{synchp}) using nominal parameters leads to a model S1 1.4 GHz luminosity estimate appropriate for Tycho,
$\nu L_{\nu} \approx 4.9 \times 10^{31}~E_{FOE}/d_{3.6}^2~ \rm ergs/sec$,
or about five times smaller than the observed luminosity. 
If we wanted to achieve a better match using the 
full scaling from equation (\ref{synchp}),
we would increase the factor $\xi K_{e,p}B^{3/2}E_{FOE}/d_{3.6}^4$ by this factor of five.

Additional information comes from the X-ray and $\gamma$-ray portions of the SED. 
Tycho's measured $10$ keV X-ray flux is $\nu F_{\nu} \approx 3.2 \times 10^{-11}~ \rm ergs/cm^2/sec$ \citep{allen99}.  
This translates to an isotropic luminosity, $\nu L_{\nu} \approx 4.9 \times 10^{34}d_{3.6}^2~ \rm ergs/sec$. 
The model S1 flux at 10 keV would be for nominal parameters, $\nu L_{\nu/sync} \approx 10^{34}~E_{FOE}/d_{3.6}^2~\rm{erg/sec}$.
This is once again about a factor five smaller, so leads to a consistent
way to compare the model to the observations. The Very Energetic Radiation Imaging Telescope Array System (VERITAS) group recently reported a
$\gamma$-ray flux above 1 TeV \cite{acciar11}, $\nu F_{\nu} \approx 3\times 10^{-13}~\rm{ergs/cm^2/sec}$. This translates to an isotropic luminosity, $\nu L_{\nu} \approx 4.7\times 10^{32}~d_{3.6}^2\rm{erg/sec}$, consistent with the properties of
our model S1.

\section{Summary}

Using kinetic, nonlinear DSA simulations, we have calculated the energy spectrum of CR protons and electrons
accelerated at blast waves from Type Ia supernovae and 
the emission contributions of those cosmic rays during Sedov-Taylor phase.
The DSA model is based on the assumptions 
that streaming instability amplified Alfv\'en waves lead to
Bohm-like diffusion, and that particle escape at the highest 
energy is negligible in the time frame of interest.
With the CR proton injection rate from the thermal plasma, $\xi \sim 10^{-3}$, obtained here,
the strong SNR shocks become
significantly modified by the formation of a CR precursor
that pre-compresses the inflowing plasma. This both weakens the
dissipative subshock and, combined with Bohm diffusion, leads
to a CR proton spectrum that is steeper at low energies than
the strong shock test particle form, $N_p(E)dE \propto E^{-2}dE$, 
but flatter at high energies. 
In addition, the drift of scattering centers with respect to the bulk plasma 
also softens the CR spectrum and the dissipation of scattering waves reduces
the CR acceleration efficiency.
These effects become more important for stronger upstream magnetic fields 
and for a lower ISM density, because the Alfv\'en speed is $v_A\propto B/\sqrt{n_{ISM}}$.

We assume the electron to proton ratio is $K_{e/p}=10^{-4}$
that is, $n_e(p_{\rm inj}) = K_{e/p} n_p(p_{\rm inj})$ near the injection momentum.
Although the value of $K_{e/p}\sim 10^{-2}$ is implied 
by the Galactic CR flux observed at Earth, 
the smaller value is preferred for source spectra that
best fit multi-band photon spectra of several young SNRs   
\citep[\eg][]{bkv09,morlino09,zira09}.  
However, since the electrons are passive with respect to the hydrodynamics 
the results of this paper can be rescaled to any (small) $K_{e/p}$.

The thermal and nonthermal continuum emissions from the model SNRs
were computed using a radiative process code {\it Cosmicp}.
The spectra are dominated in most cases by CR electron emissions in the X-ray band and below. 
Gamma-ray emissions were dominated in most cases by $p-p$ produced $\pi^0$ decay.
For the hot, low density ISM case inverse Compton electronic emissions dominated the
TeV $\gamma$-ray band, however.
We note of course that the final comparison of hadronic versus leptonic origin 
of GeV TeV $\gamma$-rays depends on the ambient matter density and the
electron to proton ratio as $n_{ISM}/K_{e/p}$ among other parameters.

We presented simple scaling relations for the 
synchrotron, IC and $\pi^0$ decay luminosities that seem to provide
appropriate approximate means to rescale the numerical results to different
times and model parameters.
We applied these relations as simple tests to SN 1006 and Tycho's SNR demonstrating
a basic consistency and a means to explore parameter space dependencies.

\section*{Acknowledgments}
We would like to thank an anonymous referee for providing many helpful comments. PPE and TWJ were supported in this work at the University of Minnesota
by NASA grant NNG05GF57G, NSF grant Ast-0908668
and by the Minnesota Supercomputing Institute for Advanced Computational Research.
HK was supported by National Research Foundation of Korea
through grant 2010-016425.
RM was supported by National Research Foundation of Korea through
grant K20702020016-07E0200-01610.
HK and TWJ also gratefully acknowledge support and hospitality from the
Kavli Institute for Theoretical Physics during their Particle Acceleration
in Astrophysical Plasmas workshop
supported in part by the National Science Foundation under Grant No. PHY05-51164.

\label{lastpage}


\begin{thebibliography}{99}

\bibitem[\protect\citeauthoryear{Abdo et al.}{2010}]{abdo10}
Abdo A., Ackermann M., Ajello M. et al., 2010, ApJ, 710, L92

\bibitem[\protect\citeauthoryear{Acciari et al.}{2011}]{acciar11}
Acciari, V. A., Aliu, E., Arlen, T. et al., 2011, ApJ, 730, L20

\bibitem[\protect\citeauthoryear{Acero et al.} {2010}]{acero10}
Acero, F., et al., 2010, A\&A, 516, 62

\bibitem[\protect\citeauthoryear{Aharonian et al.}{2007}]{aha07}
Aharonian F. A., Akhperjanian A. G., Bazer-Bachi A. R. et al., 2007, A\&A, 464, 235

\bibitem[\protect\citeauthoryear{Allen, Gotthelf \& Petre}{1999}]{allen99}
Allen G. E., Gotthelf E. V., Petre R., 1999, Proc. 26th ICRC(Salt Lake City), 3, 480

\bibitem[\protect\citeauthoryear{Allen, Petre \& Gotthelf}{2001}]{allen01}
Allen G. E., Petre R., Gotthelf E. V., 2001, ApJ,  558, 739

\bibitem[\protect\citeauthoryear{Ballet}{2006}]{ballet06}
Ballet, J.\ 2006, Advances in Space Research, 37, 1902 

\bibitem[\protect\citeauthoryear{Bamba et al.}{2003}]{bamba03}
Bamba A., Yamazaki R., Ueno M., Koyama K., 2003, ApJ, 589, 827

\bibitem[\protect\citeauthoryear{Bamba et al.}{2006}]{bamba06}
Bamba A., Yamazaki R., Yoshida T., Terazawa T., Koyama K., 2006, Adv. Space Res., 37, 1439

\bibitem[\protect\citeauthoryear{Bamba et al.}{2008}]{bamba08}
Bamba A., et al., 2008, PASJ, 60, S153

\bibitem[\protect\citeauthoryear{Bell}{2004}]{bell04}
Bell A.R., 2004, MNRAS, 353, 550

\bibitem[\protect\citeauthoryear{Beresnyak, Jones, \& Lazarian}{2009}]{bere09}
Beresnyak A., Jones T. W., Lazarian A., 2009, ApJ. 707, 1541

\bibitem[\protect\citeauthoryear{Berezhko et al.}{1994}]{berez94} 
Berezhko E.G., Yelshin V.K., \& Ksenofontov L.T. 1994, \APh, 2, 215

\bibitem[\protect\citeauthoryear{Berezhko \& V\"{o}lk}{1997}]{berez97}
Berezhko E. G., V\"olk H. J., 1997, \APh, 7, 183

\bibitem[\protect\citeauthoryear{Berezhko, Ksenofontov \&  V\"{o}lk}{2002}]{bkv02}
Berezhko E. G., Ksenofontov L.T., V\"olk, H.~J., 2002, A\&A, 395, 943 

\bibitem[\protect\citeauthoryear{Berezhko \& V\"{o}lk}{2004}]{berez04}
Berezhko E. G., V\"olk H. J., 2004, A\&A, 427, 525

\bibitem[\protect\citeauthoryear{Berezhko \& V\"{o}lk}{2007}]{berez07}
Berezhko E. G., \& V\"o{}lk, H. J., 2007, ApJ, 661, 175

\bibitem[\protect\citeauthoryear{Berezhko, Ksenofontov \&  V\"{o}lk}{2009}]{bkv09}
Berezhko E. G., Ksenofontov L.T., V\"olk, H.~J., 2009, A\&A, 505, 169

\bibitem[\protect\citeauthoryear{Berge}{2008}]{ber07}
Berge D., et al., 2008, Proc. 30th ICRC(Merida), 2, 679

\bibitem[\protect\citeauthoryear{Blandford \& Ostriker}{1978}]{bo78}
Blandford R. D., Ostriker J. P., 1978, ApJ, 221, L29

\bibitem[\protect\citeauthoryear{Blandford \& Eichler}{1987}]{be87}
Blandford R. D., Eichler D., 1987, Phys. Rep., 154, 1

\bibitem[\protect\citeauthoryear{Blasi}{2010}]{blas10}
Blasi. P., 2010, MNRAS, 402, 2807

\bibitem[\protect\citeauthoryear{Blumenthal \& Gould}{1970}]{bg70}
Blumenthal G. R., Gould R. J., 1970, RMP, 42, 237

\bibitem[Caprioli et al.(2011)]{capri10} 
Caprioli, D., Blasi, P., \& Amato, E.\ 2011, Astroparticle Physics, 34, 447 

\bibitem[\protect\citeauthoryear{Drury et al.}{2001}]{drury01}
Drury L. O'C., Ellison D. E., Aharonian F. A. et al., 2001, Space Science Reviews, 99, 329

\bibitem[\protect\citeauthoryear{Ellison \& Cassam-Chena\"i}{2005}]{ec05}
Ellison D. C., Cassam-Chena\"i, G., 2005, ApJ, 632, 920

\bibitem[\protect\citeauthoryear{Ellison et al.}{2007}]{ellison07}
Ellison D. C., Patnaude D. J., Slane P., Blasi P., Gabici S., 2007, ApJ, 661, 879

\bibitem[\protect\citeauthoryear{Funk et al.}{2007}]{funk07}
Funk S., Hinton J. A., Moriguchi Y. et al., 2007, A\&A, 470, 249

\bibitem[\protect\citeauthoryear{Gaisser}{2005}]{gai05}
Gaisser T. K., 2005, Neutrinos and Explosive Events in the Universe, Springer, p. 3

\bibitem[\protect\citeauthoryear{Garmire et al.}{2003}]{garm03}
Garmire G. P., Bautz M. W., Ford P. G., Nousek J. A., Ricker G. R. Jr, 2003, Proc. SPIE, 4851, 28

\bibitem[\protect\citeauthoryear{Gieseler, Jones, \& Kang}{2000}]{gies00}
Gieseler U. D. J., Jones T. W., Kang H., 2000, A\&A, 364, 911

\bibitem[\protect\citeauthoryear{Gonz$\rm \acute{a}${}lez}{2008}]{gon07}
Gonz$\rm \acute{a}${}lez M. M., 2008, Proc. 30th ICRC(Merida), 3, 1563

\bibitem[\protect\citeauthoryear{Hillas}{2005}]{hill05}
Hillas A. M., 2005, Journal of Physics G, 31, R95

\bibitem[\protect\citeauthoryear{Hillas}{2006}]{hill06}
Hillas A. M., 2006, arXiv:astro-ph/0607109

\bibitem[\protect\citeauthoryear{Jackson}{1999}]{jackson99}
Jackson J. D., 1999, Classical Electrodynamics, 3rd Ed., John Wiley \& Sons Inc. 

\bibitem[\protect\citeauthoryear{Jones}{1993}]{jon93}
Jones T.~W., 1993, ApJ, 619, 619

\bibitem[\protect\citeauthoryear{Kang, Jones, \& Gieseler}{2002}]{kjg02}
Kang H., Jones T. W., Gieseler U. D. J., 2002, ApJ, 579, 337

\bibitem[\protect\citeauthoryear{Kang \& Jones}{2006}]{kj06}
Kang H., Jones, T. W. 2006, \APh, 25, 246 

\bibitem[\protect\citeauthoryear{Kang}{2006}]{kang06}
Kang H., 2006, Journal of Korean Astronomical Society, 39, 95 

\bibitem[\protect\citeauthoryear{Kang et al.}{2009}]{krj09}
Kang, H., Ryu, D. \& Jones, T. W., 2009, ApJ, 695, 1273

\bibitem[\protect\citeauthoryear{Katsuda et al.}{2009}]{katsu09}
Katsuda, S., Petre, R., Long, K. S., Reynolds, S. P., Winkler, P. F., Mori, K. \& Tsunemi, H., 2009, Ap.J., 692, L105

\bibitem[\protect\citeauthoryear{Kelner \& Aharonian}{2008}]{kelner08}
Kelner S. R., Aharonian F. A., 2008, Phys. Rev. D, 78, 034013

\bibitem[\protect\citeauthoryear{Kelner, Aharonian, \& Bugayov}{2006}]{kelner06}
Kelner S. R., Aharonian F. A., Bugayov, V. V., 2006, Phys. Rev. D, 74, 034018

\bibitem[\protect\citeauthoryear{Lagage \& Cesarsky}{1983}]{lc83}
Lagage P. O., Cesarsky C. J., 1983, A\&A, 118, 223

\bibitem[\protect\citeauthoryear{Lemoine-Goumard}{2007}]{lg07}
Lemoine-Goumard M., Aharonian F. A., Berge D., Degrange B., Hauser D., Komin N., Reimer O., Schwanke U., 2007, Ap\&SS, 309, 379L

\bibitem[\protect\citeauthoryear{Lucek \& Bell}{2000}]{lucek00}
Lucek S. G., Bell A. R., 2000, MNRAS, 314, 65

\bibitem[\protect\citeauthoryear{Markiewicz, Drury, \& V\"{o}lk}{1992}]{mark92}
Markiewicz W. J., Drury L. O'C., V\"olk H. J., 1992, A\&A, 236, 487

\bibitem[\protect\citeauthoryear{Malkov \& V\"ok}{1998}]{malvol98}
Malkov M.A., V\"olk H. J., 1998, Adv. Space Res., 21, 551

\bibitem[\protect\citeauthoryear{Malkov \& Drury} {2001}]{maldru01}
Malkov M.A., Drury, L.O'C., 2001, Rep. Progr. Phys., 64, 429

\bibitem[\protect\citeauthoryear{McKenzie \& V\"{o}lk}{1982}]{mckenzi82}
McKenzie J. F., V\"olk H. J., 1982, A\&A, 116, 191

\bibitem[\protect\citeauthoryear{Morlino, Amato \& Blasi}{2009}]{morlino09}
Morlino G., Amato E., Blasi P., 2009, MNRAS, 392, 240

\bibitem[\protect\citeauthoryear{Parizot et al.}{2006}]{pari06}
Parizot E., Marcowith A., Ballet J., Gallant Y. A., 2006, A\&A, 453, 387

\bibitem[\protect\citeauthoryear{Pfrommer \& En\ss{}lin}{2003}]{pfr03}
Pfrommer C., En\ss{}lin T. A., 2003, A\&A, 407, L73

\bibitem[\protect\citeauthoryear{Ptuskin et al.}{2010}]{pzs10}
Ptuskin, V. S., Zirakashvili, V. N., Seo, E.S. 2010, ApJ 718, 31

\bibitem[\protect\citeauthoryear{Reynolds \& Ellison}{1992}]{reyn92}
Reynolds S. P., Ellison D. C., 1992, ApJ, 399, L75

\bibitem[\protect\citeauthoryear{Reynolds \& Keohane}{1999}]{reyn99}
Reynolds S. P., Keohane, J. W., 1999, ApJ, 525, 368

\bibitem[\protect\citeauthoryear{Reynolds}{2008}]{reynolds08}
Reynolds S. P., 2008, ARA\&A, 46, 89

\bibitem[\protect\citeauthoryear{Schlickeiser}{2002}]{sch02}
Schlickeiser R., 2002, Cosmic Ray Astrophysics, Springer

\bibitem[\protect\citeauthoryear{Skilling}{1975}]{skill75}
Skilling J., 1975, MNRAS, 172, 557

\bibitem[\protect\citeauthoryear{Uchiyama et al.}{2007}]{uchi07}
Uchiyama Y., Aharonian F. A., Tanaka T., Takahashi T., Maeda, Y., 2007, Nature, 449, 576

\bibitem[\protect\citeauthoryear{V\"{o}lk, Berezhko, \& Ksenofontov}{2005}]{vbk05}
V\"o{}lk H. J., Berezhko E. G., Ksenofontov L. T., 2005, A\&A, 433, 229

\bibitem[\protect\citeauthoryear{V\"{o}lk, Berezhko, \& Ksenofontov}{2008}]{vbk08}
V\"o{}lk H. J., Berezhko E. G., Ksenofontov L. T., 2008, A\&A, 483, 529

\bibitem[\protect\citeauthoryear{Wagner et al.}{2009}]{wagner09}
Wagner R. M., Lindfors E., Sillanp\"{a}\"{a} A., Wagner S., 2009, Proc. 2009 Fermi Symposium, eConf C0911022. Stanford, California

\bibitem[\protect\citeauthoryear{Webb, et al.}{1984}]{webb84}
Webb, G. M., Drury, L. O'C. \& Biermann, P., 1984, A\&A, 137, 185

\bibitem[\protect\citeauthoryear{Zirakashvili \& Aharonian}{2010}]{zira09}
Zirakashvili V.~N., Aharonian F.~A., 2010, ApJ, 708, 965

\end{thebibliography}
\end{document}